\documentclass[prd,preprint,nofootinbib,showpacs]{revtex4}
\usepackage{amsmath,amssymb,graphicx,subfigure}

\newcommand{\be}{\begin{equation}}
\newcommand{\ee}{\end{equation}}
\newcommand{\bea}{\begin{eqnarray}}
\newcommand{\eea}{\end{eqnarray}}
\newcommand{\N}{\mathcal{N}}
\renewcommand{\b}[1]{\bar{#1}}
\newcommand{\Del}{\nabla}
\newcommand{\del}{\partial}

\newcommand{\bj}{\bar{\jmath}}
\renewcommand{\ap}{\alpha^\prime}

\newcommand{\tr}{\mathrm{Tr}}
\newcommand{\re}{\mathrm{Re}}
\newcommand{\im}{\mathrm{Im}}
\newcommand{\comment}[1]{}

\newcommand{\half}{\frac{1}{2}}
\newcommand{\ads}{\ensuremath{\mathnormal{AdS}_4}}
\newcommand{\contract}{\, \lrcorner\,}
\newcommand{\cy}{\ensuremath{\mathnormal{CY}_3}}
\renewcommand{\t}[1]{\tilde{#1}}

\begin{document}
\preprint{hep-th/0507202}
\preprint{CALT-68-2567}
\preprint{UK-05-05}

\title{AdS Strings with Torsion: Non-complex Heterotic Compactifications}

\author{Andrew R. Frey}
\email{frey@theory.caltech.edu}
\affiliation{California Institute of Technology\\
Mail Code 452-48\\
Pasadena, CA 91125, USA}

\author{Matthew Lippert}
\email{lippert@pa.uky.edu}
\affiliation{University of Kentucky\\ 
Lexington, KY 40506, USA \\
and \\
University of Louisville\\
Louisville, KY 40292, USA}


\begin{abstract}

Combining the effects of fluxes and gaugino condensation in heterotic
supergravity, we use a ten-dimensional approach to find a new class of
four-dimensional supersymmetric \ads\ compactifications on
almost-Hermitian  manifolds of $SU(3)$ structure.  Computation of the
torsion allows a classification of the internal geometry, which for a
particular combination of fluxes and condensate, is nearly K\"ahler.
We argue that all moduli are fixed, and we 
show that the K\"ahler potential and superpotential proposed in the
literature yield the correct \ads\ radius.  
In the nearly K\"ahler case, we are able to 
solve the $H$ Bianchi identity using a nonstandard embedding.  Finally, 
we point out subtleties in deriving the effective
superpotential and understanding the heterotic supergravity in the 
presence of a gaugino condensate.

\end{abstract}

\date{\today}

\maketitle

\section{Introduction}\label{s:intro}

\comment{ 1. Moduli problem}

Understanding how to fix the expectation values of moduli in
compactifications of string theory has driven much of the progress 
made in discovering and categorizing string vacua over the last
few years.  The key points are that supergravity form flux
can provide potentials for moduli and that compactification manifolds other
than the traditional Calabi-Yau 3-fold (\cy) simply have fewer 
moduli (for topological reasons).  At tree-level, these ``flux vacua'' 
are relatively tractable.  For example, in some type IIB backgrounds, 
fluxes simply warp the \cy,
fix the complex structure moduli, and generate zero vacuum energy
\cite{Giddings:2001yu}. 
Similarly, in heterotic vacua with $H$ flux, 
the compact manifold is non-K\"ahler and only
the dilaton is unfixed \cite{Strominger:1986uh, deWit:1986xg, Hull:1985zy,
Hull:1986kz, Hull:1986iu}.  In more general cases, classification 
of $G$-structures has been
employed to study the resulting geometries and particularly the topological
torsions of the compactification manifolds.  

As has been known for some time, nonperturbative effects, such as 
gaugino condensation in the 4D effective field theory, can also generate
potentials for some moduli \cite{Dine:1985rz,Derendinger:1985kk}.  
In fact, it has turned out that these
nonperturbative effects happen to fix the moduli left unfixed by flux
and torsion \cite{Kachru:2003aw}.  
In the end, the vacuum can be supersymmetric with a negative
cosmological constant; that is, they are anti-de Sitter (\ads)
compactifications.  Unfortunately, however, these \ads\ compactifications,
which have been studied in type II and heterotic string theory,
are only well-understood in the 4D effective field theory.  A more
complete picture of the 10D physics would be useful, for example, in
resolving some disputes over the proper interpretation of effective field
theory in these vacua 
\cite{deAlwis:2003sn,Buchel:2003js,deAlwis:2004qh,deAlwis:2005tf}.

\comment{ 2. What we do}

In this paper, we investigate the 10D effects of gaugino condensation, 
focusing on heterotic flux
compactifications because gauge degrees of freedom are conveniently
described in the bulk supergravity.  
In particular, we will see the backreaction of the
both the flux and the condensate, making use of $G$-structures to
describe supersymmetric backgrounds.  
To our knowledge, this paper is the first use
of $G$-structures to study supergravity gaugino condensation.
Therefore, we will be relatively explicit in our calculations.  We
find that gaugino condensation is consistent with supersymmetry in
$\ads\times X^6$ compactifications where $X^6$ has $SU(3)$ structure and
belongs to a certain class of almost-Hermitian manifolds.  We find our
backgrounds to be consistent with the 4D superpotential combining the effects
of flux and gaugino condensation given in \cite{Cardoso:2003sp}, although
we raise some questions about the derivation of the superpotential.  
For specific choices of fluxes and condensate, our
compactification specializes to become nearly K\"ahler (NK), which leads
to many simplifications.  In the NK case, we are able to use a nonstandard 
embedding to solve the $H$ flux Bianchi identity, which in turn
fixes the compactification and \ads\ scales.

\comment{ 3. Organization}

Our paper is organized as follows.
We begin in section \ref{s:review} with a brief overview of flux vacua
in type II and heterotic string theories, 
including both nonperturbative effects and
their description in terms of $G$-structures.  The review provides points
of comparison to our work.  Then, in section \ref{s:SUSY}
we use the heterotic supersymmetry variation equations to derive
relations for the geometry in terms of general fluxes and gaugino
condensates.  We then use these relations to compute the components of the
torsion and classify the compactification
manifold.  Section \ref{s:potential} deals with the four-dimensional
effective theory, discussing moduli fixing and the 10D uplift of the 
gaugino condensate.  
We also verify that the proposed superpotential and K\"ahler potential
yield the correct vacuum energy.  
We do find some subtleties related to the derivation of the superpotential,
which we describe in some detail.
Specializing our flux and condensate choices in section \ref{s:NK}
yields a NK compactification.  After discussing the NK geometry, we
revisit the now simpler and explicit gaugino condensate and
moduli fixing and, in addition, solve 
the Bianchi identity.  We end in
section \ref{s:discussion} with discussion and directions of future
interest.

\textbf{Note Added:} During the final preparation of this work,
\cite{deCarlos:2005kh} appeared, which also discusses heterotic 
compactifications with torsion, flux, and gaugino condensation. We work 
from a more ten-dimensional point of view. 

\section{Review}\label{s:review}

We begin by briefly reviewing the effects of flux and gaugino
condensation in string theory compactifications.  We will particularly
highlight the parallels between the stories in 
the IIB and heterotic supergravities.

\comment{ 1. Fluxes}

The inclusion of background fluxes in string compactifications is an
old idea \cite{Strominger:1986uh, deWit:1986xg, Hull:1985zy,
Hull:1986kz, Hull:1986iu} which has more recently been applied very
successfully to the moduli problem (see \cite{Kachru:2003aw,
Curio:2005ew, Balasubramanian:2005zx, Denef:2005mm, DeWolfe:2005uu}
for some of the most recent examples).  Quantized 
Neveu-Schwarz--Neveu-Schwarz (NS) and Ramond--Ramond (RR) fluxes can
be wrapped on cycles of the compactification manifold, inducing
potentials for many moduli and deforming the background geometry.  One
can consider small additions of such fields perturbatively, ignoring
their gravitational effects (e.g. \cite{Polchinski:1995sm,
Taylor:1999ii}) and keeping the simpler \cy\ compactification.  However,
the full backreaction modifies the internal space in complicated but
interesting ways so that it is no longer \cy.

\comment{ A. IIB type B flux vacua}

In type IIB, \cite{Giddings:2001yu} showed that, under some constraints
on the brane content, the compactification manifold remains conformally
\cy, even in the presence of NS and RR flux.  This relatively mild backreaction
makes the dimensional reduction somewhat easier to analyze.  In the 4D
$\N=1$ effective theory, the fluxes induce a superpotential
of the form \cite{Gukov:1999ya} \be
\label{IIBW} W \sim \int _M G \wedge \Omega \ee where $G$, a
combination of the RR and NS fluxes and axion-dilaton, is coupled to
the geometry through the holomorphic 3-form $\Omega$.  The equations of
motion require $G$ to be imaginary self-dual (supersymmetry further requires
$G$ to be $(2,1)$), and the associated
effective potential generically fixes the dilaton and all the complex
structure moduli.\footnote{Because the required calculations are
prohibitively laborious, general arguments for fixing complex
structure moduli are typically invoked.  However, in
\cite{Denef:2005mm} the fixing was demonstrated explicitly for an
example with only three complex structure moduli.}  However, the
K\"ahler moduli, and in particular the volume modulus, remain unfixed,
resulting in a no-scale Minkowski compactification.  See
\cite{Frey:2003tf} for a review.

\comment{ B. Heterotic flux vacua}

Starting with these self-dual flux solutions in IIB, $T$ dualizing twice
and then $S$ dualizing leads to solutions of either IIB or heterotic
string theory (depending on the initial brane content) with only
NS flux \cite{Dasgupta:1999ss,Greene:2000gh,Becker:2002sx}.  
Similarly, solutions
have been studied in type IIA \cite{Cardoso:2002hd, Dall'Agata:2003ir,
Behrndt:2004km,Lust:2004ig,House:2005yc} and M-theory 
\cite{Curio:2000dw,Gauntlett:2002sc,Curio:2003ur,
Behrndt:2003zg, Kaste:2003zd, Lukas:2004ip, Behrndt:2004bh,Franzen:2005ve,
Misra:2005eu} where
fluxes also generate torsion.  While some connections between these
many flux vacua are known \cite{Becker:2002sx, Becker:2003yv,
Kachru:2002sk, Gurrieri:2002wz, Fidanza:2003zi, Grana:2004bg}, the
chain of $U$ dualities relating them remains to be worked out in
entirety.

In the case of heterotic string theory, fluxes more dramatically
affect the compactification geometry.  Adding NS flux $H$ generates
torsion in addition to warping, as implied by the title of
\cite{Strominger:1986uh}.  Imposing supersymmetry relates the flux to 
the complex structure $J$ by
\be
\label{torsionconstraint} dJ \sim \star H \, .  
\ee 
Because $J$ is closed for K\"ahler manifolds, the resulting
supersymmetric compactification has non-K\"ahler geometry.   The 
low energy superpotential is
of the form \cite{Becker:2003yv, Cardoso:2003af} \be
 \label{hetW} W \sim \int_M (H - idJ)\wedge\Omega \, .  \ee 
As in the
IIB self-dual flux case, the
effective potential fixes the complex structure moduli.  Due
to the torsion, some of the K\"ahler moduli, including the volume
modulus, are fixed as well \cite{Becker:2003yv, Becker:2003gq}.  From
another point of view, rather than fixing moduli, the compactification
geometry consistent with flux just does not have as many moduli.  
Here, the analog of the IIB volume modulus is the dilaton, which is absent
from (\ref{hetW}) and remains unfixed.

\comment{ 2. G-Structure}

Supersymmetry of a non-K\"ahler compactification
requires, rather than a special holonomy group, a reduced structure group
for the tangent bundle.  This occurs because 
the supersymmetry transformations yield 
a spinor invariant under the torsionful connection, which in
turn defines a $G$-structure.  Equivalently, the spinor bilinears define a
set of global, non-vanshing, $G$-invariant tensors.  
For $G \subset SO(n)$, these tensors include a
metric $g$ and an oriented volume $\epsilon$.   In addition, $G \subset
U(m)$ means the manifold is equipped with an almost-Hermitian metric,
an almost complex structure (ACS) $J$, and a holomorphic $m$-form $\Omega$.
As we will discuss in more detail in section \ref{ss:difftorsion}, the 
torsion is reflected in the
derivatives of the invariant tensors.
Six-dimensional non-K\"ahler manifolds with $SU(m)$ structure have
been discussed in heterotic \cite{Cardoso:2002hd,
Gurrieri:2004dt, Micu:2004tz}, types IIA \cite{Cardoso:2002hd,
Dall'Agata:2003ir, Behrndt:2004km,Lust:2004ig,House:2005yc}, 
and IIB \cite{Gurrieri:2002iw,
Frey:2003sd,Frey:2004rn,Behrndt:2005bv} strings.  Compactifications of
M-theory using $G_2$ structures  and $SU(m)$ structures
\cite{Curio:2000dw,Gauntlett:2002sc,Curio:2003ur,
Behrndt:2003zg,Kaste:2003zd,Lukas:2004ip,Behrndt:2004bh,
Franzen:2005ve,Misra:2005eu} 
have also been studied.

\comment{ 3. Gaugino Condensation and NP Effects}

The fact that the superpotentials (\ref{IIBW},\ref{hetW}) leave some
moduli unfixed led \cite{Kachru:2003aw} to reconsider nonperturbative
potentials for the other moduli, following the arguments of
\cite{Dine:1985rz,Derendinger:1985kk}.
In pure 4D $\N=1$ non-abelian gauge theory, the
gauge field $F$ becomes strongly coupled at an energy scale $\Lambda$
and the gaugino condenses: 
\be
\label{condensate} \langle \chi \chi \rangle \sim \Lambda^3 \sim
M_{\textnormal{\scriptsize UV}}^3 \, 
e^{-1/b g_{\textnormal{\tiny YM}}^2}\ , \ee 
where 
$M_{\textnormal{\scriptsize UV}}$ is
the UV cutoff scale, $b$ is an $\mathcal{O}(1)$ one-loop determinant,
and $g_{\textnormal{\scriptsize YM}}$ 
is the 4D gauge coupling.  This condensate induces an
effective superpotential of the form \cite{Affleck:1983mk}
\be
\label{gauginoW} W \sim e^{-1/b g_{\textnormal{\tiny YM}}^2} \, .  \ee

In the self-dual flux IIB vacua described above, \cite{Kachru:2003aw}
showed that the superpotentials (\ref{IIBW}) and (\ref{gauginoW})
can, in combination, freeze the remaining K\"ahler moduli, in particular
the volume modulus.  Specifically, if the gaugino condensation
occurs in a D7-brane gauge group, the gauge coupling depends on the
(fixed) dilaton and the volume modulus.  Therefore, the volume 
modulus is fixed as well.\footnote{Instantonic D3-branes can 
introduce a similar nonperturbative superpotential, and leading 
order $\ap$ corrections
additionally restore dependence of the potential on the volume modulus
\cite{Becker:2002nn}.}
Most relevantly for us, however, the combined flux and gaugino superpotentials
lead to supersymmetric \ads\ vacua
\cite{Kachru:2003aw, Denef:2004dm, Denef:2005mm}.
In addition, metastable $\mathnormal{dS}_4$ minima have been
constructed by adding branes \cite{Kachru:2003aw} or otherwise
breaking supersymmetry \cite{Burgess:2003ic, Brustein:2004xn} to
increase the vacuum energy.

In a heterotic \cy\ compactification, gaugino condensation on its own 
drives the 
4D dilaton to the strong coupling region.  However, when the condensate
is balanced against
$H$ flux, the 4D dilaton is finite, and the vacuum is a no-scale
Minkowski spacetime with broken supersymmetry
\cite{Dine:1985rz, Derendinger:1985kk}.  Even though the 
cosmological constant vanishes, the superpotential $W\neq 0$.  This solution
is nontrivial, as the wrapped fluxes are quantized and fractional
values are needed to match the exponentially small contribution of the
condensate \cite{Gukov:2003cy}.\footnote{In \cite{Gukov:2003cy},
one-loop corrections were used to fix the volume modulus.  Alternatively,
worldsheet instantons, as with the IIB D3-instantons, can give a potential
to K\"ahler moduli.}  From a 10D perspective, $SU(3)$ holonomy of the \cy\
requires that the flux and the condensate be $(3,0)+(0,3)$
forms.  More recently, \ads\ and even de Sitter vacua have
been shown to arise from gaugino condensation in the strong string coupling
limit (heterotic M-theory) 
\cite{Buchbinder:2003pi,Becker:2004gw,Buchbinder:2004im}.  Finally, 
\cite{Cardoso:2003sp} has, analogously to the IIB case, considered adding
gaugino condensation to compactifications with both $H$ flux and torsion.

\section{Conditions for Supersymmetric \ads } \label{s:SUSY}

In this section, we will examine the conditions for supersymmetry in
heterotic supergravity in the presence of both $H$ flux and gaugino
condensation.  We will find that gaugino condensation is actually
consistent with supersymmetry in $\ads\times X^6$ compactifications.
In particular, we will study the backreaction of the condensate, using
the $G$-structure formalism.  Since this is the first use of
$G$-structures to study supergravity gaugino condensation, our
calculations will be relatively explicit.

As it turns out, the $H$ flux and gaugino condensate induce a
significant backreaction on the geometry of the compact manifold
$X^6$.  In the well-studied
\cite{Hull:1985zy,Hull:1986kz,Hull:1986iu,Strominger:1986uh,
Becker:2002sx,  Becker:2003yv, Becker:2003gq, Becker:2003sh,
Cardoso:2003af} case of supersymmetric Minkowski compactifications,
the $H$ flux generates a torsion and the internal manifold ceases to
be K\"ahler.   We will see a similar
but even more dramatic effect in the \ads\ case here.

\subsection{Ansatz and $SU(3)$ Structure}\label{ss:structure}

To set our field conventions, we give here the string frame action of
the effective supergravity for the bosonic fields and gaugino
\cite{Bergshoeff:1989de}\footnote{Compared with
\cite{Bergshoeff:1989de},  we have made the following rescalings:
$\chi_{BdR} = \sqrt{2} \chi$,  $\psi_{BdR} = \sqrt{2} \psi$, $H_{BdR}
= H/3\sqrt{2} $, and  $\phi_{BdR} = e^{2\phi/3}$.}  \be
\label{action} S = \frac{1}{2\kappa_{10}^2} \int d^{10}x \sqrt{-g}
e^{-2\phi} \left\{ R + 4\del_M \phi \del^M \phi - \frac{1}{2} \left| H
-\frac{1}{2} \Sigma \right|^2-
\frac{\kappa_{10}^2}{2g_{10}^2}\tr\left( F^2  +4\b\chi\Gamma^M
D_M\chi\right)\right\}\ ,  \ee  where \be \Sigma_{MNP} =
\frac{\kappa_{10}^2}{g_{10}^2} tr \b\chi \Gamma_{MNP} \chi \ee  is the
gaugino condensate, $2\kappa_{10}^2 = (2\pi)^7{\ap}^4$, and
$\kappa_{10}^2/g_{10}^2 = \ap/4$ (see \cite{Polchinski:1998rr}).
Index, form, and spinor conventions are given in appendix
\ref{a:conventions}.  For convenience, we will define $T = H- \half
\Sigma$.   We will work in string frame with a metric ansatz of the
form \be
\label{metricansatz} ds^2 = e^{2A}\hat{g}_{\mu\nu} dx^\mu dx^\nu +
g_{mn} dx^m dx^n \ee where $\hat{g}_{\mu\nu}$ is an \ads\ metric of
radius $R$ and $g_{mn}$ is the metric on the internal space $X^6$.  We
will at times factor out the volume modulus by writing $g_{mn} =
e^{2u} \tilde{g}_{mn}$, where the volume of $X^6$ is $V_6 = e^{6u}
(2\pi\sqrt{\ap})^6$ and $\tilde{g}_{mn}$ is a  fiducial metric  with
volume $\tilde V_6 = (2\pi\sqrt{\ap})^6 $.  In the interest of
preserving $SO(3,2)$ symmetry, we take the other fields neither to
depend on nor have components in the \ads\ directions.

The string-frame supersymmetric variations of the dilatino $\lambda$,
gaugino $\chi$, and gravitino  $\psi_M$ are \bea
\label{dilatinovar} \delta\lambda &=& -\half \del_M \phi \Gamma^M
\varepsilon +  \frac{1}{24} H_{MNP}\Gamma^{MNP} \varepsilon +
\frac{1}{96}\Sigma_{MNP}\Gamma^{MNP} \varepsilon \\
\label{gauginovar} \delta\chi &=& -\frac{1}{4}
F_{MN}\Gamma^{MN}\varepsilon \\
\label{gravitinovar} \delta\psi_M &=& \Del_M\varepsilon - \frac{1}{8}
H_{MNP}\Gamma^{NP}\varepsilon  +\frac{1}{96}\Sigma_{NPQ}\Gamma^{NPQ}
\Gamma_M \varepsilon \, .\eea  The supersymmetry parameter  $\varepsilon$
is a 10D Majorana-Weyl spinor with positive chirality which we
decompose into 4D and 6D positive chirality Weyl spinors, $\zeta$
(anticommuting) and $\eta$ (commuting), as \be
\label{spinordecomp} \varepsilon = e^{A/2} \left(\zeta \otimes \eta +
\zeta^* \otimes  \eta^* \right) \ , \ee where the warp factor is
included for convenience of normalization.  Because $\varepsilon$ is
Majorana, there is only one independent  positive chirality 6D spinor,
meaning the solution has an $SU(3)$ structure.  We will see later that
we can normalize $\eta^\dagger\eta \equiv 1$.

With that normalization, we can define the $SU(3)$ structure in terms
of \be
\label{su3def} J_m{}^n = -i\eta^\dagger \gamma_m{}^n \eta \ ,\ \
\Omega_{mnp} = \eta^T \gamma_{mnp} \eta \, .  \ee By the Fierz
identities for $\eta$, $J$ is an almost complex structure ($J^2=-1$),
and (with $J$ written as a form) \be J \wedge \Omega = 0 \ ,\ \
\Omega \wedge \bar\Omega = -\frac{4i}{3} J\wedge J\wedge J \ .  \ee
The $SU(3)$ structure determines the almost Hermitian metric $g_{mn}$
and volume form  $\epsilon = (i/8) \Omega \wedge \bar\Omega$.
$\Omega$ is $(3,0)$ with  respect to $J$ and imaginary self-dual.
See, for example,  \cite{Gauntlett:2003cy} for a review of
$SU(n)$-structures.  Henceforth, if we  count holomorphic and
antiholomorphic indices on a form, we do so with  respect to $J$.

\subsection{Algebraic Relations}\label{ss:algebraic}

In supersymmetric vacua,  the supersymmetric variations
(\ref{dilatinovar} - \ref{gravitinovar}) are all required to vanish.
We start by inserting (\ref{spinordecomp}) into the $\mu$ component of
(\ref{gravitinovar}), noting that the covariant derivative is 
\bea
\label{Delepsilon} \Del_\mu \varepsilon &=& \hat\Del_\mu \varepsilon +
\half \Gamma_\mu\,^N\del_N A \varepsilon \nonumber\\  
&=& e^{A/2}\left[\gamma_\mu \zeta\otimes\left(-m \eta^*+\half \del_n A
\gamma^n\eta\right)+\gamma_\mu\zeta^*\otimes\left( -\b m \eta -\half
\del_n A \gamma^n \eta^*\right)\right] \ ,
\eea 
where $m$ is
proportional to the 4D superpotential and is related to the \ads\
radius $R$ and cosmological constant $\Lambda$ by  
\be
\label{m} |m| = \frac{1}{2R} = \half \sqrt{\frac{|\Lambda|}{3}} 
\ee  
(in 4D string frame).  The second equality of (\ref{Delepsilon}) follows
from the \ads\ covariant derivative $\hat\Del_\mu \zeta = -\b
m\gamma_\mu\zeta^*$ for a Weyl Killing spinor (see
\cite{Behrndt:2003zg,Behrndt:2004bh}).  Subdividing
(\ref{gravitinovar}) by 4D (or 6D) chirality gives  \be
\label{one} \frac{1}{96} \Sigma_{mnp} \gamma^{mnp} \eta = \half \del_m
A \gamma^m \eta  - m \eta^* \, .  \ee Similarly, the dilatino
variation (\ref{dilatinovar}), combined with (\ref{one}), becomes \be
\label{four} \frac{1}{24} T_{mnp}\gamma^{mnp}\eta = \half \del_m
(\phi-3A)\gamma^m\eta  + 3m\eta^*\ .  \ee

Now we can simplify in terms of the $SU(3)$ structure.   Using
(\ref{one}), its adjoint, and  some gamma matrix algebra, we compute
\be
\label{SigmaJ1} \frac{1}{16} \Sigma_{mnp}J^{np}  = \frac{1}{96}
\Sigma_{npq} \eta^\dagger \left\{ \gamma_m, \gamma^{npq} \right\} \eta
= \del_n A J_m\,^n \, .  \ee In terms of our form notation
(\ref{contraction}), this is \be\label{SigmaJ} J\contract \Sigma = -8
dA\contract J\ .\ee Furthermore, multiplying (\ref{one}) by $\eta^T$,
we find 
\be
\label{SigmaOmega} \Omega\contract\Sigma = -16 m \, .  \ee 
Similarly,
using (\ref{four}) we can compute 
\bea
\label{TJOmega} J\contract T &=& 2d(3A-\phi)\contract J \\
\Omega\contract T &=&12m\ .  
\eea
Using
(\ref{decompose}), we have altogether 
\bea 
\Sigma&=&-2\left(\b
m\Omega+m\b\Omega\right)+\Sigma_0-4J\wedge\left(dA \contract
J\right)\nonumber\\ 
T&=& \frac{3}{2}\left(\b
m\Omega+m\b\Omega\right)+T_0+J\wedge\left( d(3A-\phi)\contract
J\right)\ ,\label{SigmaTdecomp} 
\eea where
$J\contract\Sigma_0=\Omega\contract\Sigma_0=0$ and similarly for
$T_0$.  Finally, the gaugino variation gives conditions which are
familiar from  Calabi-Yau compactifications \cite{Candelas:1985en}:
\be\label{duyand11} F\contract \Omega=J\contract F=0\ .  \ee With
respect to the ACS, these conditions are respectively $F=F_{(1,1)}$
and the Donaldson-Uhlenbeck-Yau (DUY) equation.

We should note that we have not imposed that $\Sigma=a\Omega + \b
a\bar\Omega$  as in the usual CY compactification.   In particular,
equation (\ref{SigmaJ}) is not identically  zero.  Instead, our
condensate is intentionally general at this  stage, and we will
further address the dimensional reduction of the  gaugino in section
\ref{ss:gauginocondensate}.

\subsection{Differential Relations and Torsion}\label{ss:difftorsion}

We return now to the gravitino variation.  Multiplying (\ref{one}) by
$\gamma_q$ and rearranging using some gamma  matrix algebra, we get \be
\label{two} \frac{1}{96} \Sigma_{mnp} \gamma^{mnp} \gamma_q \eta =
\frac{1}{16}  \Sigma_{qmn}\gamma^{mn}\eta -\half \del_m A \eta - \half
\del_m A \gamma_q\,^m  \eta  + m \gamma_q \eta^* \, .  \ee Then we can
write the internal components of (\ref{gravitinovar}) as \be
\label{three} \Del_m\eta = \half \del_n A \gamma_m{}^n\eta +
\frac{1}{8} T_{mnp}\gamma^{np}  \eta- m \gamma_m \eta^* \, .   \ee
Note that (\ref{three}) implies\footnote{The last term does not
contribute  due to the antisymmetry of $\gamma_m$.} that we can set
$\eta^\dagger \eta =1$ constant.  Therefore, the $SU(3)$ structure is
properly normalized.

While the current form of (\ref{three}) is useful for our
calculations, it is also instructive to see that $\eta$ is parallel
with respect to a torsionful connection.   Inserting $\eta^\dagger
\eta=1$, we can use the Fierz identity (\ref{fierz2}) along with  some
gamma matrix algebra and the anti-self-duality of $\bar\Omega$ to
write  
\be\label{omegaeta} 
\bar\Omega_{mpq} \gamma^{pq}\eta =8\gamma_m\eta^* \ \
\textnormal{and similarly}\ \ \Omega_{mpq} \gamma^{pq} \eta = 0\ .
\ee 
Inserting these into (\ref{three}), we find   \be
\label{etawithtorsion} \Del_m\eta = \half \del_n A \gamma_m{}^n\eta +
\frac{1}{8} \tau_{mnp}\gamma^{np} \eta\, ,  \ee where the (intrinsic)
torsion is  \be
\label{tau} \tau^m{}_{np} = T^m{}_{np} - \b m\Omega^m{}_{np} -
m\bar\Omega^m{}_{np}\ .   \ee

It is also important to interpret the spinor equations  (\ref{three})
and (\ref{etawithtorsion}), starting with (\ref{three}).  Although the
last term seems unusual, (\ref{three}) is actually just a  Killing
spinor equation for a Weyl spinor, much like the Killing equation for
a spinor in \ads.  Here, though, we have a connection with contorsion
$\kappa_{pmn}=2g_{m[p}\del_{n]}A-\frac{1}{2}T_{pmn}$.    To write
(\ref{three}) in a canonical form, shift the contorsion into the
covariant derivative and define
$\eta'=e^{i\beta/2-i\pi/4}\eta+e^{-i\beta/2+i\pi/4}\eta^*$ where
$m=e^{i\beta}|m|$.  Then we have $\Del_m\eta' = -i|m|\gamma_m \eta'$,
the defining equation for a real Killing spinor
\cite{Lichnerowicz:1987rx}.  On the other hand, (\ref{etawithtorsion})
shows that $\eta$ is parallel with respect to an alternative connection, one
with contorsion
$\kappa_{pmn}=2g_{m[p}\del_{n]}A-\frac{1}{2}\tau_{pmn}$.  In the more
familiar \cy\ compactification, the Killing spinor is parallel with
respect to the Levi-Civita connection, meaning that the manifold has
$SU(3)$ holonomy, not just $SU(3)$ structure.

To understand the relation between $SU(3)$ holonomy and structure, it
is useful to examine the torsion in more detail.\footnote{See
appendix \ref{a:torsions} for the definition  of the torsion compared
to the contorsion.}  What is important is not the contorsion itself
(in fact, it is possible to remove the warp factor term from the
contorsion by rescaling the internal metric to $g_{mn}=e^{2A}\b
g_{mn}$) but the intrinsic torsion, or simply the torsion.   The
torsion $\tau$ cannot be removed by any conformal rescaling, and it is
actually a topological quantity. The torsion gives a topological
obstruction to special holonomy (of the Levi-Civita connection) for
$X^6$.   Rather, after metric rescaling, it is the connection with 
torsion $\tau$ 
which has $SU(3)$ holonomy. We can see this obstruction
directly in the $SU(3)$ structure.  The torsion, thought of as an
$su(3)$ valued one-form, can be decomposed into five irreducible
$SU(3)$ modules $W_i$, which give  $dJ$ and $d\Omega$ as follows: \bea
\label{dJtorsion} dJ &=& \frac{3i}{4} \left( W_1 \bar\Omega -
\bar{W_1}\Omega \right) + W_3 + J\wedge W_4 \\
\label{dOmegatorsion} d\Omega &=& W_1J\wedge J + J\wedge W_2 +
\Omega\wedge W_5  \, .  \eea Decomposing the torsion with respect to
the ACS, $dJ$ has $(3,0)$ and $(0,3)$ parts which
give $W_1$ and $\bar{W}_1$ and a $(2,1)+(1,2)$ part which can further
be decomposed into a primitive part $W_3$ and a nonprimitive part
$J\wedge W_4$.  Similarly, $d\Omega$ has a $(3,1)$ part $W_5$ and a
primitive $(2,2)$ part $J\wedge W_2$.  The non-primitive $(2,2)$
piece of $d\Omega$ redundantly gives $W_1$.  Additionally, $3W_4-2W_5$ is
conformally invariant.

What, then, are the derivatives of the $SU(3)$ structure tensors 
in our compactification?
Using (\ref{su3def},\ref{etawithtorsion}), we find that 
\bea  
\Del_m J_n\,^p &=& -\del_q A \left(g_{mn}J^{pq} - g^{pq}J_{mn} - \delta_m^p
J_n\,^q +  \delta_n^q  J_m\,^p \right)\nonumber\\  
&& + \frac{1}{2}
\tau_{mn}\,^q J_q\,^p  - \frac{1}{2}\tau_{mq}\,^p J_n\,^q \, ,\label{Jwithtau} 
\eea  
showing that $J$ is covariantly constant with the
appropriate contorsion (once again, the $\del A$ terms can be removed
by rescaling the metric). Since $\tau$ is the $SU(3)$ holonomy
torsion, this is consistent with the fact that $J$ is an $SU(3)$
singlet.

For computing $dJ$ there is a more useful expression of $\Del J$, 
which we find starting from (\ref{three}).  After some
gamma matrix algebra, we arrive at  
\bea
\Del_m J_n\,^p &=& -\del_q A
\left(g_{mn}J^{pq} - g^{pq}J_{mn} - \delta_m^p J_n\,^q + \delta_n^q
J_m\,^p \right)\nonumber\\  && + \half T_{mn}\,^q J_q\,^p - \half
T_{mq}\,^p J_n\,^q - 2\, \im(\b m\Omega_{mn}\,^p)\ .
\label{seven}
\eea 
We can relate (\ref{seven}) to (\ref{Jwithtau}) using the Fierz
identity  (\ref{fierz1}), which yields \be
\label{omegaJ}  \Omega_{mn}\,^{p} = \frac{i}{2} \Omega_{mn}\,^{q}
J_q\,^p- \frac{i}{2}  \Omega_{mq}\,^{p} J_n\,^q \,.   \ee (In fact,
deriving (\ref{omegaJ}) is one step in showing that $\Omega$ is
$(3,0)$ with respect to $J$.)  Continuing, we antisymmetrize to find
\be
\label{dJcalc} (dJ)_{mnp} = 6 \del_{[m}A J_{np]} - 3T_{[mq}\,^q
J_{p]q} - 6\,  \im(\b m\Omega_{mnp}) \, .   \ee  
To substitute in for
$T_{[mq}\,^q J_{p]q}$, we use (\ref{four}) to write  
\bea  -3 T_{[mn}{}^q J_{p]q}  &=& \frac{-i}{12} T_{mnp}\eta^\dagger
\left[\gamma_{mnp},\gamma^{qrs} \right]  \eta + \frac{1}{6} T^{qrs}
\epsilon_{mnpqrs} \nonumber\\ 
&=& 6 \del_{[m}(\phi-3A) J_{np]} +
(\star T)_{mnp} + 12\, \im(\b m \Omega_{mnp})\ .   \eea  
Plugging into
(\ref{dJcalc}), we find   \be dJ = 2 d(\phi-2A) \wedge J + \star T +
6\, \im(\b m\Omega) \, .   \ee  Taking the Hodge dual of $T$ from
(\ref{SigmaTdecomp}),  \be  \star T = \frac{3i}{2} {\bar m} \Omega -
\frac{3i}{2} m \bar\Omega +  d(3A-\phi) \wedge J + \star T_0\ .  \ee
The expression for $dJ$ becomes  
\be
\label{dJ} dJ = \frac{3i}{2}(m\bar\Omega - \bar{m}\Omega)+ \star T_0 +
d(\phi-A) \wedge J   \, .   \ee  
Reading directly from (\ref{dJ}),
$W_1 = 2m$, $W_4 = d(\phi -A)$, and $W_3=\star T_0$.

We can perform a similar calculation to find $d\Omega$.   Some gamma
matrix algebra yields \be (d\Omega)_{mnpq} = 3 (dA \wedge
\Omega)_{mnpq} + 6T_{m[s}\,^s\Omega_{pq]s}  + 8m \re(\eta^\dagger
\gamma_{mnpq}\eta)  \, .  \ee Again, we use (\ref{four}) to substitute
in for $T_{m[s}\,^s\Omega_{pq]s}$ as  \be 6 T_{m[n}\,^s\Omega_{pq]s} =
2 \left(d(\phi-3A)\wedge\Omega\right)_{mnpq} - 6m\, \re(\eta^\dagger
\gamma_{mnpq} \eta) \, .  \ee Then \be
\label{dOmega} d\Omega =  d(2\phi-3A) \wedge \Omega + 2m J\wedge J \,
.  \ee It is easy to confirm the result for $W_1$, and we see $W_5 = -
d(2\phi-3A)$.   Finally, we can conclude that $W_2 = 0$, since there
is no  primitive piece of $d\Omega$.

To summarize, we have found the torsion classes of $X^6$ to be \bea
W_1 &=& 2m\, , \ W_2 = 0\, ,\ W_3 = \star T_0 \nonumber \\ W_4 &=&
d(\phi-A)\, \textnormal{, and} \ W_5 = - d(2\phi-3A) \,
.\label{torsions} \eea

\subsection{Almost Hermitian Geometry}\label{ss:geometry}

Manifolds with $U(m)$ structure,  called almost-Hermitian manifolds,
were classified \cite{Gray:1980} into sixteen categories depending on
which components $W_1$ to $W_4$ of the torsion are non-zero.  For a
manifold to be complex,  both $W_1=0$ and $W_2=0$ are required.
Simply put, on a complex manifold, $dJ$ should have no $(3,0)+(0,3)$
parts, and $d\Omega$ should have no $(2,2)$ parts (just by counting
holomorphic indices).   Some examples of almost-hermitian manifolds
are given in Table \ref{t:manifolds} along with their non-zero torsion
components and whether or nor they are complex.

\begin{table}[t]
\begin{center}
\begin{tabular}{|c|ccccc|c|} \hline  Manifold Name & $W_1$ & $W_2$  &
$W_3$  &  $W_4$  &  $W_5$  & Complex?\\ \hline \hline    Hermitian & 0
& 0 & $W_3$ & $W_4$ & $W_5$ & Y \\ Balanced & 0 & 0  & $W_3$ & 0 &
$W_5$ & Y \\ Special Hermitian & 0 & 0  & $W_3$ & 0 & 0 & Y \\
K\"ahler & 0 & 0  & 0 & 0 & $W_5$ & Y \\  Calabi-Yau & 0 & 0  & 0 & 0
& 0 & Y \\  Nearly K\"ahler & $W_1$ & 0  & 0 & 0 & 0 & N \\  Almost
K\"ahler & 0 & $W_2$ & 0 & 0 & 0 & N \\  Quasi-K\"ahler & $W_1$ &
$W_2$ & 0 & 0 & 0 & N \\  Half-flat & $\im W_1$& $\im W_2$&
$W_3$&0&0&N\\ Semi-K\"ahler & $W_1$ & $W_2$ & $W_3$ & 0 & 0 & N \\
$\mathcal{G}_1$ & $W_1$ & 0  & $W_3$ & $W_4$ & $W_5$ & N \\
$\mathcal{G}_2$ & 0 & $W_2$  & $W_3$ & $W_4$ & $W_5$ & N \\ \hline
\end{tabular}
\end{center}
\caption{\label{t:manifolds}Almost hermitian manifolds are  classified
by torsion.}
\end{table}

It may be helpful to place some familiar examples in this context.  If
the first four torsion components are all zero,  the manifold is
K\"ahler.  The vanishing of $W_5$ in addition signals that the
manifold is Calabi-Yau.  The non-K\"ahler compactifications considered
in  \cite{Hull:1985zy,Hull:1986kz,Hull:1986iu,Strominger:1986uh,
Becker:2002sx,  Becker:2003yv,
Becker:2003gq,Becker:2003sh,Cardoso:2003af} are in fact Hermitian with
$W_5=-2W_4$, and they are conformally balanced
\cite{Cardoso:2002hd,Gauntlett:2003cy}.  Half-flat manifolds have also
been of recent interest
\cite{Chiossi:2002,Gurrieri:2002wz,Gurrieri:2002iw,Gurrieri:2004dt,
Lust:2004ig,Chiossi:2004ig}.

In our $\ads \times X^6$ examples, we have found that generically
$X^6$ has non-zero $W_1$, $W_3$, $W_4$, and $W_5$. The internal
manifold $X^6$ is therefore not complex, but is of the type
$\mathcal{G}_1$ \cite{Gray:1980}; however, there are some interesting
special cases.  Let us start by taking $m=|m|e^{i\beta}$ and
redefining $\Omega=ie^{i\beta}\Omega'$, so $W_1$ is purely imaginary
when (\ref{dJtorsion},\ref{dOmegatorsion}) are written in terms of
$\Omega'$.  Then, for $J\contract T=0$, $W_4=(2/3)W_5=d\phi$, so $X^6$
is conformally half-flat with $W_2=0$.  If we additionally have
$J\contract\Sigma=0$, $W_4=W_5=0$, and $X^6$ is 
actually half-flat.\footnote{Note that this is a 
different class of half-flat manifolds than that
considered in \cite{Lust:2004ig}.}  Under the further condition  that
the primitive part of $T$ (and therefore $\star T$) vanishes, then
$X^6$ is conformally nearly K\"ahler (for $J\contract\Sigma\neq 0$) or
actually nearly K\"ahler (NK) ($J\contract\Sigma = 0$).  We will
return in some detail to the case of NK $X^6$ in section \ref{s:NK}.

\section{The 4D Effective Theory} \label{s:potential}

In this section, we will relate the supersymmetric backgrounds
discussed above to the 4D effective field theory that arises from the
compactification.  First, we give a dictionary relating parameters of the
10D and 4D descriptions, naturally leading to a discussion of moduli
fixing. Then we raise more subtle issues regarding 
the gaugino condensate.  Finally, we give a consistency check for 
proposed K\"ahler and superpotentials, finding a puzzle for the
superpotential.

\subsection{Dictionary and Moduli Fixing}\label{ss:modulifixing}

To understand the interplay between the 4D and 10D physics in our
\ads\ backgrounds, we should begin by understanding the dimensional
reduction of the fields.  Along the way, we will see how different
moduli are fixed in different manners.  

First off, the Einstein frame is defined with respect to the 4D part of
the string frame metric by $g_{E,\mu\nu}=e^{6u-2\phi}g_{\mu\nu}$,
and the Planck mass is $m_P^2=1/\pi\ap$ (see
appendix \ref{a:normalization}).  Note that we are including the 
moduli expectation values in the rescaling, so they do not enter in the
Plank mass.  Then, converting to the Einstein frame, the action
(\ref{action}) for the gauge fields reduces to 
\be\label{Faction}
S_F = -\frac{1}{4g_{\textnormal{\tiny YM}}^2}\int d^4x\, \sqrt{-g_E}\,
\tr F_{\mu\nu}F_E^{\mu\nu}\ ,\ \ \frac{1}{g_{\textnormal{\tiny YM}}^2}
=\frac{1}{4\pi}e^{6u-2\phi}\ ,
\ee
where the subscript $E$ on the gauge field denotes that the spacetime 
indices have been raised with the Einstein frame metric.\footnote{We are
implicitly assuming that the moduli are frozen at their expectation
values.  Further, for simplicity, we are assuming that the dilaton and 
warp factor are constant over the compactification.  This assumption does
not actually change the results qualitatively.}  
The gauge theory is weakly coupled in the regime of large
radius and/or weak coupling.

The gaugino decomposes under dimensional reduction as 
\be\label{chi10}
\chi = \left(\chi_6 \otimes \chi_4 + \chi_6^* \otimes \chi_4^*\right)
+\chi_{KK}\ , 
\ee
where $\chi_4$ carries all the gauge indices and $\chi_6$ satisfies a
zero mode equation (other polarizations are superpartners of other 
polarizations of the gauge field and are lumped with higher Kaluza-Klein
modes in $\chi_{KK}$).  Using a Majorana spin flip identity (and dropping
KK modes),
we can reduce the gaugino action from (\ref{action}) to 
\be\label{chiaction}
S_\chi  = -\frac{2}{g_{\textnormal{\tiny YM}}^2} \int d^4x\,\sqrt{-g_E}\,
\tr\b\chi_E \gamma^\mu_E D_{E,\mu}\chi_E\ ,\ \ 
\chi_E= \exp\left[\frac{3\phi-9u}{2}\right]\chi_4\ .
\ee
The important point is that the canonically normalized gaugino in Einstein
frame is rescaled from the dimensionally reduced gaugino given in 
(\ref{chi10}).  In fact, the rescaling is precisely the appropriate
rescaling for a field of dimension $m^{3/2}$.\footnote{Note that because 
it is a connection, the gauge field is not rescaled.}  Therefore, if 
$\langle \overline{\chi_E^*}\chi_E\rangle \equiv \Lambda_E^3$ sets the scale of 
gaugino condensation in Einstein frame, we see that mass rescaling 
simply tells us that $\Lambda^3=\langle\overline{\chi_4^*}\chi_4\rangle$ is the 
equivalent mass scale in the string frame.
From (\ref{chi10}), we can therefore write
\bea
\Sigma_{mnp}&=&\frac{\kappa_{10}^2}{g_{10}^2}\,\left[
\langle\tr\overline{\chi^*_4}\chi_4\rangle
\left(\chi_6^T\gamma_{mnp}\chi_6\right)
-\langle\tr\b\chi_4\chi^*_4\rangle
\left(\chi_6^\dagger\gamma_{mnp}\chi^*_6\right)
\right]+\Sigma'_{mnp}\nonumber\\
&=&\frac{\ap}{4} \left[\Lambda^3\left(\chi_6^T\gamma_{mnp}\chi_6\right)
-\b\Lambda^3\left(\chi_6^\dagger\gamma_{mnp}\chi^*_6\right)\right]
+\Sigma'_{mnp}\ .\label{sigmareduce}  \eea
This is the relation between the 4D and 10D descriptions of the 
condensate ($\Sigma'$ is the expectation value of other Kaluza-Klein 
modes, presumably generated by some other quantum effect).  
For now, let us assume that
a significant portion of $\chi_6^T\gamma_{mnp}\chi_6$ lies 
along $\Omega$; then we have $\ap \Lambda^3\sim |m|$.  The effective
field theory description of gaugino condensation is consistent as long
as the condensate scale is less than the Kaluza-Klein scale,
$\Lambda\ll m_{KK}=e^{-u}/\sqrt{\ap}$, which we find is valid as long
as 
\be\label{validity}3u+\frac{1}{2}\ln \left(\ap|m|^2 \right)\leq 0\ .\ee
Fortnately, (\ref{validity}) is satisfied in the large radius regime of
the compactification as long as the \ads\ radius is sufficiently large.

Now we are in position to see how the 4D dilaton modulus $S$
and the radial modulus $T$ are fixed, which we understand from the point
of view of the effective field theory.\footnote{The moduli are defined 
in appendix \ref{a:normalization}.}  
Analogously to \cy\ compactifications 
\cite{Dine:1985rz}, the superpotential becomes
\be\label{superdil}
W\sim C(T)+ Ae^{ia S}\ ,
\ee
The effective potential freezes the VEV of the dilaton.  
Similarly, because of the torsion $dJ$ in the superpotential (\ref{hetW}),
we expect $C$ to depend on $T$, so the supergravity is no longer 
no-scale, and $T$ will be fixed.  Any complex-structure-like moduli
will presumably be fixed by additional dependencies in $C$ and $A$.

Simultaneously, in our supersymmetric vacuum, the vacuum value of $W$ 
determines the (negative) cosmological constant.  In the string frame, 
as we mentioned in (\ref{m}), the superpotential therefore determines
$m$, which is a derived value, like a modulus. 
To evaluate $m$, we note that in a supersymmetric \ads\ vacuum 
all contributions to the superpotential should be of similar magnitude, which 
agrees with our 10D results (\ref{SigmaTdecomp},\ref{dJ}).
In fact, (\ref{SigmaOmega},\ref{SigmaTdecomp}) set the condensate
proportional to the \ads\ scale $m$.  Combining the expressions for the
condensate scale gives an approximate equation giving the $\phi$ in
terms of $m$ and $u$.  We find  
\be\label{mlambda}
m\sim \ap \Lambda^3\sim \frac{e^{-3u} e^{iaS}}{\sqrt{\ap}}\ ,
\ee
using the Kaluza-Klein mass as the ultraviolet cutoff of equation 
(\ref{condensate}).
Assuming that we can obtain the small coupling regime in the gauge
theory, the \ads\
is large and nearly flat, with a very tiny cosmological constant.
Additionally, (\ref{mlambda}) implies that the effective field theory
approximation is valid (see (\ref{validity})).

Before we reinterpret some of the above in 10D language, we should 
also give a caveat.  In checking the validity of the effective theory
and in estimating the \ads\ curvature, we have assumed that 
$\chi_6^T\gamma_{mnp}\chi_6\sim \Omega_{mnp}$, which is true for
\cy\ compactifications.  However, as we will discuss in further detail
in \ref{ss:gauginocondensate} below, in these non-K\"ahler examples 
the zero mode $\chi_6$ may lead to
other components.

We can also understand how most of the moduli are fixed from a 10D 
perspective.  Starting with complex-structure-like moduli, we note that
$X^6$ is not complex, and the torsion can greatly reduce the number of 
deformations of the almost
complex structure.  One argument that the remaining moduli are fixed
is similar to that of \cite{Rohm:1985jv} for \cy\ compactifications with
flux and condensates:
As in the \cy\ case, $H=A \Omega+\b A\b\Omega +\cdots$.  
On the \cy, the $\cdots$ vanish, and the flux $H$
takes only quantized values \cite{Rohm:1985jv}.  The complex structure
moduli are therefore fixed to discrete values that allow $H$ to lie
in the integral cohomology.  In our vacua, $dH\neq 0$, so the precise
quantization condition is not known.  However, we expect that there
will be some quantization mechanism, which will again force $\Omega$ to
take a compatible form and freeze the complex-structure-like moduli.

For the volume modulus $u$, the situation
is much as in the case of vanishing condensate, in which 
the radial modulus is fixed by flux
via the torsion constraint (\ref{torsionconstraint}).  In our \ads\
case, (\ref{torsionconstraint}) is generalized to (\ref{dJ}).
Now $u$ is fixed by the presence of non-zero $W_3$
as can be seen by the following scaling argument:  Under the dilation
$u \to u+c$ and $m \to e^{-c} \, m$,  
\be \label{scaling} 
g_{mn} \to e^{2c} g_{mn} \ ,\ \ J_{mn} \to
e^{2c} J_{mn} \ ,\ \ \Omega_{mnp} \to e^{3c}\Omega_{mnp}\ .   
\ee
However, we do not expect that the primitive $T_0$, and 
therefore $W_3$, should scale under the
dilation; for example, if it is closed, $H_0$ should be an element of 
the integral cohomology.  Therefore, under the scaling,   
$dJ \to e^{2c}\left\{\frac{3i}{2}(m\bar\Omega -
\bar{m}\Omega) + d(\phi-A) \wedge J \right\} + (\star T)_0 \neq
e^{2c} dJ$.  The rescaled manifold is not a solution,
and $u$ is not a modulus; from the perspective of the 4D theory, it 
must develop a potential (possibly a very steep one).  In fact, a similar
scaling argument has been given in the absence of a gaugino condensate, 
see \cite{Becker:2002sx,Becker:2003yv}.

The $H$ flux definition in terms of Chern-Simons forms (or equivalently
its Bianchi identity) gives an alternate and comprehensive way
to understand moduli fixing from the 10D perspective. As discussed in 
\cite{Becker:2003yv,Becker:2003sh}, $H$ appears as torsion in the 
Lorentz-Chern-Simons term, so $H$ is defined only implicitly.  The key
point is that solving for $H$ also constrains $u$.  In fact, in the
most general case, there should be enough components of $H$ to constrain
additional complex-structure-like moduli, as well.  Finally, because
$H$ has components proportional to $m$, we expect that $m$ is constrained.
In section \ref{ss:NKBianchi}, we will show that, for the special case of a 
nearly K\"ahler compactification, the Bianchi identity simplifies 
considerably and can be solved using a nonstandard embedding to give an 
explicit value for $m$.

\subsection{Gaugino Condensate}\label{ss:gauginocondensate}

We remind the reader that $\Sigma=a\Omega+\b a\b\Omega$ when we
consider gaugino condensation in Calabi-Yau compactifications.
However, we have so far allowed a more general form of $\Sigma$ for
two reasons.  One is that we wish to leave open the possibility that
10D quantum effects may turn on different components of the gaugino
condensate than the 4D effective theory.  We will not say anything
concrete about this possibility, but we return to this point in
section \ref{ss:future}.  The other reason that we have left $\Sigma$
general is that we do not expect the 4D condensate to lift to
$\Sigma=a\Omega+\b a\b\Omega$ in a compactification with torsion.  We
will explain why here.

Let us begin by briefly considering the dimensional reduction of the
gaugino.  
The Kaluza-Klein zero-modes (given by $\chi_6$ in (\ref{chi10})) 
satisfy the Dirac equation
following from the  action (\ref{action}) 
\be\label{zerodirac}
\left(\gamma^m D_m +\del_m(2A-\phi)\gamma^m
-\frac{1}{24}T_{mnp}\gamma^{mnp} \right)\chi=0\ , \ee 
where we use the
full spinor $\chi$ 
to include the full gauge structure.    However,
because the unbroken low-energy gauge group commutes with the
background gauge fields $A_m$, the (adjoint) gaugino zero modes are
neutral  under the gauge background.  Therefore, we can replace
$D_m\to\Del_m$ and $\chi\to\chi_6$ in (\ref{zerodirac}).

It is straightfoward to see that $\chi_6=\eta$, the supersymmetry
parameter, in the case of a \cy\ compactification.  Simply put, on a
\cy\ (with or without gaugino condensate and compensating $H$-flux),
the Dirac equation is $\gamma^m\Del_m\chi_6=0$, while $\eta$ is the
unique covariantly constant spinor.  Clearly, then, with the same
normalization, $\chi_6=\eta$.  In that case, (\ref{sigmareduce}) gives
$\Sigma=a\Omega+\b a\b\Omega$.

On the other hand, in our solutions, the gaugino Dirac equation
(\ref{zerodirac}) is generally not the same as the Dirac equation
following from the supersymmetry equation (\ref{etawithtorsion}).  For
example, if there is any primitive part of $T$, it appears in
different proportions in the two Dirac equations, as do the warp
factor and dilaton.  In other words, $\chi_6$ is not a singlet of the
$SU(3)$ structure.  So, in general, we do not expect the low-energy
condensate to lift to $\Sigma=a\Omega+\b a\b \Omega$.

A very interesting question, then, is whether the background $\Sigma$
is always the same as the uplift of the 4D gaugino condensate, or
whether some other 10D quantum effects are responsible for some
components of the background condensate $\Sigma$.  That is, can
$\Sigma$ be identified with the condensate of the effective theory in
our backgrounds?  While we cannot find the gaugino zero mode in all
backgrounds, we will see in section \ref{ss:NKgaugino} that the answer
is ``yes'' in some backgrounds, while the effective theory cannot
account for all of the condensate in other backgrounds.

\subsection{A Check of Potentials}\label{ss:potentials}

Now we will present a consistency check of proposals for both the
superpotential and K\"ahler potential.  
The appropriate generalization of the Gukov-Vafa-Witten superpotential
for the heterotic theory has been argued \cite{Cardoso:2003af,Cardoso:2003sp,
Becker:2003gq} to include both the contributions of torsion and gluino
condensation.  
Consistent with those calculations, we employ the ansatz  
\be \label{superpotential} 
W = \frac{m_p^3}{\sqrt{4\pi}} \frac{1}{(2\pi\sqrt{\ap})^5}\int \left( H + b
\, idJ + c \Sigma \right)   \wedge \tilde\Omega \, . \ee  
The 3-form $\t\Omega = e^{-3u}\Omega$ 
corresponds to the fiducial metric $\t g_{mn}$.\footnote{See appendix 
\ref{a:normalization}, equations 
(\ref{gvwsuper},\ref{kahlerpot}), for normalizations.}  To agree with
\cite{Cardoso:2003af,Cardoso:2003sp,Becker:2003gq} in our conventions,
we should take $b=+1$.

The K\"ahler potential for a \cy\ compactification is given by  
\be \label{kahlerpotential} 
\mathcal{K}=-3\ln(-i(T-\b T)) -\ln(-i(S-\b S))
-\ln\left(\frac{i}{(2\pi\sqrt{\ap})^6} \int\t\Omega\wedge\b{\t\Omega}\right) 
+3 \log 2 \ee 
in terms of the 4D superfields $S$ and $T$
and the rescaled holomorphic 3-form $\tilde\Omega$.\footnote{The 
4D superfields are defined in appendix \ref{a:normalization}.}
In terms of 10D variables, the K\"ahler potential is simply 
\be\label{kahler} 
\mathcal{K} =  2\phi - 12 u - 4 \log 2 \, .  \ee
Unfortunately, the moduli spaces of non-K\"ahler manifolds, even those with
$SU(3)$ structure, are unknown, and
(\ref{kahlerpotential}) appears inapplicable.  However, as argued in
\cite{Gurrieri:2004dt,Micu:2004tz}, 
we can perhaps think of $X^6$ as a deformation of a
Calabi-Yau for which (\ref{kahlerpotential}) is valid (this assertion
is supported by direct calculation on half-flat manifolds
\cite{Gurrieri:2004dt}).  Alternately,
\cite{Micu:2004tz} further notes that for manifolds with a single
K\"ahler modulus, the K\"ahler potential is sufficiently simple as to
be universal.  We will adopt (\ref{kahlerpotential}) as an ansatz
whose consistency is verified by the calculation of this section.
For simplicity's sake, we are ignoring any possible effect from a warp
factor.

The 4D effective potential in Einstein frame 
is now given by (see (\ref{sugraV})) 
\be  V = m_P^{-2} e^{\mathcal{K}}
\left( \sum_{a,b} \mathcal{K}^{a\b b} D_a W \bar D_{\b b} \b W  
- 3 |W|^2 \right)
\ee  
where the sum is over all moduli $a$ and $b$ and $\mathcal{K}_{a{\bar b}} =
\del_a\bar\del_{\bar b}\mathcal{K}$.  Because we are considering a
supersymmetric vacuum, the K\"ahler covariant derivatives vanish,
leaving only the final term to give the cosmological constant.  
As a first check,
we can confirm that the proposed potentials give the correct cosmological
constant for our background.

Plugging from equations (\ref{SigmaTdecomp},\ref{dJ}) into the
superpotential and using 
$\b\Omega \wedge \tilde\Omega=8i e^{3u}\tilde \epsilon$, we find 
immediately  
\be \label{super} W = i \frac{m_P^3}{\sqrt{4\pi}} 
(2\pi\sqrt{\ap}) (4-12b-16c) e^{3u} m \, .  \ee
Putting everything together, the absolute Einstein frame cosmological constant 
is 
\be\label{lambdaE}
\Lambda_E = -\frac{V}{m_P^2} = 3\pi m_P^2\ap e^{2\phi-6u}
|1-3b-4c|^2 |m|^2 = \frac{1}{4}e^{2\phi-6u}|1-3b-4c|^2\Lambda_s
\ee
after using (\ref{m}) and (\ref{einsteinaction}), where $\Lambda_s$ is the
(absolute) cosmological constant in string frame. The exponential of
the moduli is precisely the necessary combination for conversion of a 
mass dimension 2 constant between frames, so the K\"ahler potential and
superpotential are consistent if
\be\label{Wcondition}
1-3b-4c=\pm 2\ .\ee
Some consistent solutions are
\be\label{Wsols}
b=1,\  c=-1;\ \
b=1,\  c=0;\ \
b=-1,\  c=\frac{1}{2};\ \
b=-1,\  c=\frac{3}{2}\ .
\ee

We can also check the superpotential and K\"ahler potential 
by computing the supersymmetry variation of the Einstein frame 
gravitino.  For our metric ansatz (\ref{metricansatz}), the gravitino
with the canonical supersymmetry transformation in Einstein frame is
\be\label{einsteingravitino}
\psi_{E,\mu} \equiv \Psi_\mu\otimes\eta+\Psi_\mu^*\otimes\eta^*
= e^{(3u-\phi))/2} \left[ \psi_\mu +\Gamma_\mu
\left( \frac{1}{2} \Gamma^m\psi_m +\lambda\right)\right]\ .\ee
Here $\Gamma^M$ is still the 10D string frame Dirac matrix, and
$\Psi$ and $\Psi^*$ are the positive and negative chirality spinors in 4D.
This linear combination is the same one which diagonalizes the gravitino
kinetic term.  After some algebra, we find
\be\label{einsteinvariation}
\delta\Psi_\mu = e^{(3u-\phi)/2} \left[\Del_\mu\zeta -\frac{1}{48}
H_{mnp}\b\Omega^{mnp}\gamma_\mu\zeta^* -\frac{1}{2}\eta^\dagger
\gamma^m(\Del_m\eta^*)\gamma_\mu\zeta^*\right]\ .\ee
Note that any contributions from $\Sigma$ have \textit{canceled} between
the external gravitino variation and the dilatino variation.
Using (\ref{omegaeta}), we can rewrite the last term as
\be\label{etadeleta}
-\frac{1}{2}\eta^\dagger\gamma^m(\Del_m\eta^*) = 
-\frac{1}{16}\eta^T\gamma_{np}\Del_m\eta^*\b\Omega^{mnp}
= \frac{i}{48}(dJ)_{mnp}\b\Omega^{mnp} +\frac{1}{16}(\Del_m\eta^T)
\gamma_{np}\eta^*\b\Omega^{mnp}\ ,\ee
and the last term vanishes, again because of (\ref{omegaeta}).
In the end, then, 
\be\label{einsteinvariation2}
\delta\Psi_\mu = e^{(3u-\phi)/2} \left[\Del_\mu\zeta -\frac{1}{48}
\left(H-idJ\right)_{mnp}\b\Omega^{mnp}\gamma_\mu\zeta^*\right]\ .\ee
Again, all variables on the right hand side of (\ref{einsteinvariation2})
are given in the 10D string frame.  From \cite{Cremmer:1982en}, we know
that the second term in the gravitino variation must be proportional
to $e^{\mathcal{K}/2}\b W$, so we find (up to overall phase)
\be\label{finalW}
W = \frac{m_p^3}{\sqrt{4\pi}} \frac{1}{(2\pi\sqrt{\ap})^5}\int \left( H + idJ
\right)\wedge\tilde \Omega\ ,\ee
consistent with the second solution of (\ref{Wsols}).

Oddly, the superpotential derived in this way does not contain the 
gaugino condensate $\Sigma$, in contrast to the proposal of 
\cite{Cardoso:2003sp}.  Our result is especially counterintuitive because
the effective 4D field theory certainly gains a nonperturbative 
superpotential associated with gaugino condensation.  Furthermore, 
our result (\ref{einsteinvariation2}), when evaluated in our \ads\ 
backgrounds, gives $\Del_\mu\zeta = -\b m \gamma_\mu \zeta^*$, which 
is exactly the substitution we used in equation (\ref{Delepsilon}).  It 
would therefore seem inconsistent to add a nonperturbative superpotential.
One possible resolution is that, in the ``dictionary'' between the
10D supergravity and 4D effective theory, $H$ or $dJ$ includes the 
nonperturbative part of the superpotential; we will return to this question 
in our discussion of future directions.

\section{Nearly K\"ahler Compactification} \label{s:NK}

We have found in section \ref{s:SUSY} that generically $X^6$ has
non-zero $W_1$, $W_3$, $W_4$, and $W_5$ and is of type
$\mathcal{G}_1$.  However, by imposing certain conditions on the
choice of fluxes and condensates, we can easily turn off all the
torsions but $W_1$, making the manifold nearly K\"ahler.  Nearly
K\"ahler (NK) compactifications have been of recent interest  in
massive type IIA supergravity \cite{Behrndt:2004km, Behrndt:2004mj},
M-theory \cite{Lukas:2004ip}, and heterotic strings
\cite{Micu:2004tz}.    In this section, we  will first list some
properties of  NK manifolds and relate them to our compactifications;
in following subsections, we elaborate on our earlier discussions in
the special case of an NK compactification.

\subsection{Nearly K\"ahler Geometry}\label{ss:NKgeometry}

Nearly K\"ahler manifolds, in many ways the simplest non-K\"ahler
manifolds, have been studied by Gray \cite{Gray:1970, Gray:1971,
Gray:1976} and have many interesting mathematical
properties.\footnote{For a more recent work which reviews the
properties of NK manifolds see \cite{Nagy:2004}.}  The defining
property of NK $2n$-folds is weak $SU(n)$ holonomy, which appears in
the $SU(3)$ holonomy of the torsional derivative.   Among other
curvature identities, every NK manifold in six dimensions is Einstein
and has vanishing first Chern class \cite{Gray:1976}.  In addition,
the cone over any six-dimensional NK manifold will have holonomy $G_2$
\cite{Bryant:1987,Bryant:1989,Bar:1993}.

Many examples of NK manifolds are known due to a theorem by  Gray
\cite{Gray:1972} that 3-symmetric spaces\footnote{A 3-symmetric  space
has global isometries $\theta_p$ for each point $p$, where $p$ is the
fixed point of $\theta_p$.  Each $\theta_p^3=1$ and is holomorphic
with respect to a canonically associated ACS.} have an NK metric and
ACS (or are products of such manifolds).   Furthermore,
\cite{Gray:1972} showed that 3-symmetric spaces can be identified in a
natural way with cosets of connected Lie groups and proceeded to
classify such cosets.  For example, the simplest of a NK manifold is
the sphere $S^6\simeq G_2/SU(3)$.

Another nice result in the mathematical literature is due to Grunewald
\cite{Grunewald:1990}, regarding the Killing spinor.  Any spin
manifold with a real Killing spinor, with respect to the Levi-Civita
connection, is NK, and, conversely, any NK manifold has a real Killing
spinor.   In fact, following from (\ref{tauNK}), the Killing number
is $|m|/2$.  This point is important in ensuring that supersymmetry
is truly preserved for an NK compactification with the appropriate
constraints on the flux  backgrounds (in other words, given an NK
manifold and the appropriate algebraic relations for the flux and
condensate, there does exist an appropriate supersymmetry transformation
parameter on the manifold).  Additionally,
\cite{Lichnerowicz:1987rx} has shown that manifolds with real Killing
spinors are compact.

Let us now remind the reader of the appearance of NK manifolds in our
supersymmetric backgrounds.  In general, the supersymmetric vacua
described in section \ref{ss:difftorsion} are not NK because $W_3$,
$W_4$, and $W_5$ are nonvanishing.  However, as mentioned in section
\ref{ss:geometry}, we can set the primitive part $T_0=0$ (implying
$H_0=\Sigma_0/2$) to remove the $W_3$ torsion.  Then choosing the
nonprimitive $J\contract T=0$, we find that $X^6$ is conformally NK
(meaning that $W_4=(2/3)W_5$).  Finally, requiring
$J\contract\Sigma=0$ also forces $W_4=W_5=0$, so we have a truly
nearly K\"ahler manifold.  In the following, we will distinguish two
cases: $\Sigma_0=0$, which we will simply call NK, and $\Sigma_0\neq
0$, which we will denote NK$^\prime$.  It is useful also to simplify (for
either case) 
\bea 
H &=& \frac{1}{2}(\b m\Omega+m\b\Omega)
\label{HNK}\\ 
\Sigma&=& -2(\b m\Omega+m\b\Omega)\ \left\{\, +\Sigma_0
\textnormal{ for NK}^\prime \frac{}{}\right\}\label{SigmaNK}\\ 
\tau&=&
\frac{1}{2}\left(\b m\Omega+m\b\Omega\right)\ .\label{tauNK}
\eea

\subsection{Gaugino Condensation for NK}\label{ss:NKgaugino}

At the end of section \ref{ss:gauginocondensate}, we asked if the
background $\Sigma$ can be identified with the gaugino condensate of
the effective field theory.  Although we do not know the general
answer, we will show here that such an identification is consistent in
NK compactifications.

Since $dA=d\phi=0$ in the NK case, the gaugino zero mode Dirac
equation (\ref{zerodirac}) becomes the covariant Dirac equation for
the connection with torsion $\tau_\chi=T/3=(\b m\Omega+m\b\Omega)/2$
with the specific form of $T$ given by (\ref{HNK}, \ref{SigmaNK}).  
But the supersymmetry parameter $\eta$
is constant with respect to the $SU(3)$ structure torsion
(\ref{tauNK}), which is the same!  Therefore, the gaugino zero mode is
$\chi_6=\eta$.  Then the uplifted 4D condensate (\ref{sigmareduce}) is
just $\Sigma = \kappa_{10}^2/g_{10}^2(\langle
\overline{\chi_4^*}\chi_4 \rangle\Omega+\langle
\b\chi_4\chi_4^*\rangle\b\Omega)$,  which is consistent with the 10D
decomposition for NK compactifications.  In other words, $\Sigma$ can
be generated completely by the condensate in the effective  4D theory.
However, in the NK$'$ case, the additional background condensate
$\Sigma_0$ is not of the correct form and must therefore be generated
by some as yet unknown quantum effect.

\subsection{Moduli fixing for NK}

In the NK case, the moduli fixing is slightly different from the
general case described in section \ref{ss:modulifixing}.  Because $W_3
= 0$, it would seem that $u$ is now unfixed.  However, we can use a
scaling argument instead to fix the size of $X^6$ in terms of the \ads\
scale $m$.  For a dimensionless NK manifold of unit radius, the almost
complex structure $J_0$ and $SU(3)$ structure $\Omega_0$ are related
by \cite{Cvetic:2002kj}: 
\be\label{dJOmegaforNK} dJ_0 = \frac{3i}{2}( \bar\Omega_0 - \Omega_0) \ ,\
d\Omega_0 = 2 J_0\wedge J_0 \, .  \ee
We have already computed the derivatives of $J$ (\ref{dJ}) and
$\Omega$ (\ref{dOmega}), which in the NK case reduce to
\be\label{dJOmegaforNK2}
dJ = \frac{3i}{2}(m\bar\Omega - \bar m \Omega) \ ,\
d\Omega = 2m J\wedge J \, .
\ee

To compare the two formulae, we need to rescale the unit radius NK 
manifold, first by giving it dimensionful coordinates.  Since our
fiducial metric has volume $(2\pi)^6\ap{}^3$, we rescale the coordinates
by $x\to 2\pi\sqrt{\ap}/\omega_6^{1/6}$, where $\omega_6$ is the volume
of the unit NK manifold.\footnote{For example, for $S^6$, 
$\omega_6=16\pi^3/15$.}
Additionally, in rescaling $\tilde g_{mn}\to g_{mn}$, we must rescale
the ACS and 3-form, so we eventually get 
\be\label{rescale}
J=\frac{4\pi^2\ap}{\omega_6^{1/3}}e^{2u} J_0\ ,\ 
\Omega = \frac{8\pi^3\ap{}^{3/2}}{\omega_6^{1/2}}e^{3u+i\beta}\Omega_0
\ee
(including a possible phase $\beta$ for the 3-form).
Equations (\ref{dJOmegaforNK},\ref{dJOmegaforNK2}) match for
\be
\label{NKm}
m = \frac{\omega_6^{1/6}}{2\pi\sqrt{\ap}} e^{-u+i\beta} \, .
\ee
So, for the NK solution, the \ads\ and compactification scales are 
essentially the same (differing by a factor of order unity).  The criterion
(\ref{validity}) for the validity of effective field theory is therefore
only satisfied in the NK case if 
\be\label{NKvalid}
u\lesssim \frac{1}{2}\ln 2\pi - \frac{1}{12}\ln\omega_6 \ ,
\ee
which means the compactification is in the large radius regime only 
if $\omega_6$ is sufficiently small.   For example, for the case of 
$S^6$, $\omega_6$ is small enough that $u\lesssim 1/2$ is valid.
The direct relation between $m$ and $u$ also means that the 
NK manifold has no radial modulus, properly speaking.  Because changing 
$m$ changes the \ads\ boundary conditions, $u$ corresponds to a 
nonnormalizable mode.

In the general case we found weakly coupled solutions with large $u$ 
and exponentially large \ads\ radius, but unfortunately, that 
is not longer possible for NK compactifications.  
Using (\ref{NKm}) in (\ref{mlambda}), 
the equation for the dilaton now becomes
\be
2u \sim -e^{6u - 2\phi}\ ,
\ee
which, in the most optimistic case, gives $u\sim 0$ and 
$g_{\textnormal{\scriptsize YM}}^2\lesssim 1$.  

Incidentally, \cite{Brustein:2004xn,Curio:2005ew} argued that 
the general superpotential ansatz (\ref{superpotential}) cannot
be responsible for fixing complex structure moduli unless
there is $(2,1)+(1,2)$ flux.  However, our
NK background should have no massless scalars.  The resolution is
that, due to the relation (\ref{dJOmegaforNK2}), there are no 
variations of $\Omega$ independent of variations of $J$.  That is to
say, NK manifolds have no ``complex structure'' moduli.  In fact,
\cite{Micu:2004tz} has already proposed that NK manifolds have
no such moduli.

As we mentioned in the general analysis of moduli fixing (section
\ref{ss:modulifixing}), the $H$ flux Bianchi identity also constrains
the moduli.  Since the NK $SU(3)$ structure is so simple, 
we can solve the Bianchi identity, which we discuss in the following section.

\subsection{Bianchi Identity}\label{ss:NKBianchi}

We have yet to consider the restrictions imposed by the Bianchi 
identity for $H$,
\be
\label{bianchi} dH = \frac{\ap}{4} \left(\tr R_-\wedge R_- - 
\frac{1}{30} \tr F\wedge F \right) \ee 
where $R_-$ is the Ricci two-form constructed
using the torsion $\tau_- = -H$.  Note that this is opposed to the
curvature $R(\tau)$ constructed with the torsion $\tau$ 
associated with the $SU(3)$ structure given in eqn (\ref{tau}).

An advantage of the NK case over our more general one (or even NK$^\prime$) 
is that the Bianchi identity becomes sufficiently tractable that we can make 
explicit computations.  In particular, a significant simplification 
results from the fact that, in 
the NK case, (\ref{HNK}) and (\ref{tauNK}) imply $\tau_- = -(m\bar\Omega + 
{\bar m}\Omega)/2 = - \tau$.  

In torsion-free \cy\ compactifications, the Bianchi identity is
typically simplified by  imposing the standard embedding of the spin
connection into the gauge connection,  canceling to two terms on the
right-hand side of (\ref{bianchi}) to yield $dH =0$ and  breaking the
gauge group from $E_8 \to E_6$ 
We could employ a similar tactic here, but since (\ref{HNK}) and
(\ref{dJOmegaforNK})  gives
\be
\label{dHforNK} dH = 2|m|^2 J\wedge J  , \ee 
setting $\tr R_-\wedge
R_- - \frac{1}{30} \tr F\wedge F$ would mean $m=0$, requiring  a \cy\
compactification with no flux or condensate and a Minkowski spacetime.
Instead we will try to embed a more general spin connection with
torsion proportional to $\tau$ into the gauge connection.

We denote by $R_\xi$ the Riemann curvature constructed using the
torsion  $\tau_\xi = \xi \tau = \xi(m\bar\Omega + \bar{m}\Omega)/2$.
Using identities  (\ref{contorsion}, \ref{riemanntorsion}), we can
relate $R_\xi$, of which  $R_- = R_{-1}$ is a special case, to
$R(\tau)=R_1$, which for consistency we will hereafter  denote $R_+$: 
\be
{R_\xi}_{mnpq} = {R_+}_{mnpq} - (\xi-1) \nabla^+_{[p} \tau_{q]mn} 
+\frac{\xi-1}{2}\tau_{mnr}\tau_{pq}{}^r
+\frac{(\xi-1)^2}{2}{\tau^r}_{n[q}\tau_{p]mr} \, .  \ee 
Since $\Omega$ is an $SU(3)$ invariant tensor, the covariant derivative
vanishes.  Simplifying the other terms
using (\ref{omegacontractions}) gives 
\bea
 {R_\xi}_{mnpq} &=& {R_+}_{mnpq} 
+\frac{1}{2}|m|^2 \left[ \frac{\xi^2-1}{4}(J\wedge J)_{mnpq}
+\frac{(\xi-1)(\xi-3)}{2}J_{mn}J_{pq}\right.\nonumber\\
&&\left.
-\frac{(\xi-1)(\xi-3)}{2}\left(g_{mp}g_{nq}-g_{mq}g_{np}\right)\right]\ .
\label{RxitoRplus}
\eea

We can then plug (\ref{RxitoRplus}) into $\tr R_\xi\wedge R_\xi$.
With some extensive algebra,
we obtain 
\be
\label{xicherntopluschern} \tr R_\xi\wedge R_\xi = \tr R_+ \wedge R_+
- 6(\xi - 1)^2 |m|^4 J\wedge J \, .  \ee
In order to simplify cross terms in the expansion of the trace, 
we use (\ref{etawithtorsion},\ref{commutators}) to deduce that  
\be\label{RJidents}
{R_+}_{mnpq} J^{pq} = 0\ ,\ \
{R_+}_{[mn}{}^{sr} (J\wedge J)_{pq]rs} =  4{R_+}_{[mnpq]}.\ee  
From
(\ref{RxitoRplus}) with $\xi=0,$ we find ${R_+}_{[mnpq]}  =
-\frac{1}{3} |m|^2 (J\wedge J)_{mnpq}$.  

Setting $\xi = -1$ in (\ref{xicherntopluschern}) enables us to
substitute  $\tr R_+\wedge R_+$ for $\tr R_-\wedge R_-$ in the Bianchi
identity (\ref{bianchi}).   A natural idea would now be to embed the
$SU(3)$ spin connection with torsion  $\tau$ into the gauge
connection.   However, setting $\frac{1}{30} \tr F\wedge F =  \tr
R_+\wedge R_+$ and employing (\ref{dHforNK}) reduces (\ref{bianchi})
to \be 2|m|^2J\wedge J  = -24 \ap |m|^4 J\wedge J \, , \ee whose only
solution is an unacceptable $m = 0$.

Instead, we use (\ref{xicherntopluschern}) again to write \be \tr
R_-\wedge R_- = \tr R_\xi\wedge R_\xi + 6(\xi-3)(\xi+1) |m|^4 J\wedge
J \, , \ee We impose a nonstandard embedding with $\frac{1}{30} \tr
F\wedge F =  \tr R_\xi\wedge R_\xi$.  Any connection with torsion
other than $\tau$ has holonomy  $SO(6)$ rather than $SU(3)$, so this
embedding breaks the gauge group  $E_8\to SO(10)$ or $SO(32)\to
SO(26)$.  The Bianchi identity (\ref{bianchi}) now  just a condition
on $m$, \be \ap |m|^2 = \frac{1}{3(\xi-3)(\xi+1)} \, ,  \ee which has
positive solutions for $|m|$ when $\xi < -1$ or $\xi > 3$.  To our
knowledge, there seems to be no reason why $\xi$ is constrained (within
the allowed region), but, once it and $m$ are chosen, changing $\xi$ would
be a nonnormalizable mode in \ads\ and therefore not a modulus.  This
is as we discussed in the previous subsection.  In a sense, the \ads\
radius determines the embedding of the spin connection into the gauge 
connection.

\section{Discussion of Open Questions}\label{s:discussion}

In this section, we will discuss open questions about supersymmetry
and gaugino condensation in the heterotic theory.

\subsection{Superpotential}\label{ss:superdiscuss}

In section \ref{ss:potentials}, we realized that the superpotential of
the 4D effective theory is still not completely understood.  We can
point out two related issues that, as yet, lack explanations.

First, our supersymmetric \ads\ backgrounds are incompatible with a
superpotential of the form $\int (T+idJ)\wedge \Omega$, even though
that form is suggested by the 10D supergravity action
\cite{Cardoso:2003sp}.  In particular, we showed in (\ref{Wsols}) that
such a superpotential leads to the wrong  value of the cosmological
constant.  Additionally, reminding ourselves of \cy\ compactifications
with gaugino condensates confirms that the superpotential cannot
depend on $H$ and $\Sigma$ only through $T$.  In those
compactifications, at least to lowest order, the vacua are no-scale
and Minkowski, as can be seen in the effective theory.  However,
supersymmetry is broken by the condensate, which implies that the
superpotential cannot vanish, $W\neq 0$.  On the other hand,  $T=0$ in
\cy\ compactifications.  So, indeed, even though $\Sigma$ enters the
supergravity only through $T$, the effective superpotential has some
alternate dependence on $\Sigma$.

More disturbing, perhaps, is our discovery that $\Sigma$ does not
seem to enter the superpotential at all, which we found by computing
the Einstein frame gravitino SUSY variation.  Certainly, we know that
the condensate generates a nonperturabative superpotential in the
effective field theory.  Perhaps this nonperturbative  superpotential
should not appear in our semi-classical treatment of the background,
even though we  explicitly left $\Sigma\neq 0$.  On the other hand,
adding any new contribution to the superpotential would conflict with
our initial choice of \ads\ Killing spinor.

What is the real story?  One possibility is that the dictionary from
10D variables to the 4D effective field theory is nontrivial in the
presence of gaugino condensates.  In other words, perhaps some of the
$H$ flux or torsion $dJ$ contains the nonperturbative part of the
superpotential.  However, we think it more likely that the condensate
affects the 10D supergravity in some more subtle way.  For example,
the gaugino kinetic term $\b\chi \gamma^M D_M\chi$ should also acquire
an expectation value when the 4D gaugino condenses.  Specifically,
making use of the decomposition (\ref{chi10}) and the   gaugino zero
mode Dirac equation (\ref{zerodirac}), we can see that
\be\label{kineticcondensate} \langle\b\chi\Gamma^M D_M\chi\rangle =
\frac{1}{24}T_{mnp}\left(  \langle\b\chi_4\chi_4^*\rangle
\chi_6^\dagger \gamma^{mnp}\chi_6^* -
\langle\overline{\chi_4^*}\chi_4\rangle\chi_6^T\gamma^{mnp}\chi_6
\right) = \frac{1}{4}T\contract\Sigma \ee for zero-momentum gaugini in
\ads.  Here, we assume that the entire condensate is generated in the
4D effective theory.  Most likely, however, understanding all the
effects of the condensate in 10D  will require understanding the 1PI
effective action of the supergravity.

\subsection{Equations of Motion and Bianchi Identity}

In section \ref{s:SUSY} we set the supersymmetry variations to zero in
order to find supersymmetric backgrounds.  However, we have not
explicitly shown that these backgrounds are solutions to the equations
of motion; generally speaking, the supersymmetry conditions do not
imply all of the equations of motion.  The independent equations of
motion must be imposed additionally to guarantee that the backgrounds
are indeed solutions.  For example, the heterotic supersymmetric
background with constant dilaton presented in \cite{Cardoso:2003sp}
was shown in \cite{Gauntlett:2003cy} not to satisfy the $H$ and $F$
equations of motion.  Similarly, in the context of massive IIA,
\cite{Lust:2004ig} argued that both the form field equations of motion
and Bianchi identity were additionally required for supersymmetric
vacua to be solutions.

In the case considered here, the gaugino condensate makes the
derivation of the equations of motion more subtle.  In particular,
the expectation value of the gaugino kinetic term should appear in the
equations of motion.  Again, it seems likely that the appropriate 10D
equations of motion are given by the 1PI effective action of the
supergravity in the presence of a condensate.

In addition to the equations of motion, prospective solutions must
also satisfy the Bianchi identity.   For the particular case of NK
compactifications  we were able to make considerable progess by
relying on a generalization of  the standard embedding.  A similar
calculation may be possible for the less  tractable general case.
However, it seems unlikely that the solution could be so simple.  In
general, we will not be able to employ this approach, and instead we
may need to involve the gauge field in some more complicated way.
Perhaps a series solutions in powers of $\ap$ is the best that we can
expect, as in  \cite{Becker:2003yv,Becker:2003sh}.

\subsection{Topology Change}

One can approach flux vacua from two different directions.  We have
taken the view that one looks for self-consistent combinations of
compactification manifold, fluxes, condensate, and \ads\ radius.  The
topological data and choice of flux will be consistent with only
discrete values of continuous moduli, and not all discrete choices
will necessarily yield consistent backgrounds.   In particular, \cy\
compactifications with $H$ flux are not supersymmetric; different
topological data, including non-K\"ahlerity (and non-complexity if
gaugino condensates are added), are required.   In this view,
therefore, we do not describe the fluxes as backreacting  on a
preexisting \cy\ geometry.

However, one could instead choose to begin with a particular flux-free
\cy\ compactification and then turn on fluxes using appropriate branes
as domain walls.  This is the context in which the backreaction of
fluxes can be made precise.  In type II supergravity,
\cite{Gurrieri:2002wz}  argued, in fact, that NS5-brane domain walls
in \cy\ compactifications are mirror symmetric to topology-changing
domain walls (presumably wrapped Kaluza-Klein monopoles), which in
fact transform a \cy\ into a half-flat manifold.

In our case, the reader might think that topology change could occur
via instantons, allowing the decay of nonsupersymmetric Minkowski
vacua into our \ads\ vacua.  However, including gravitational effects,
tunneling could only occur if the Minkowski vacuum were lifted by loop
or string effects to de Sitter, and the end state would be a big
crunch  universe rather than \ads\ \cite{Coleman:1980aw}.
Nonetheless, it would be interesting to trace the connection between
the Minkowski and \ads\ vacua.

\subsection{Dualities}\label{ss:dualities}

Another possible angle to explore is the existence of an \ads/CFT$_3$
duality.  Beyond the standard AdS/CFT duality with compact spheres
(e.g.\ $AdS_5 \times S^5$) \cite{Aharony:1999ti}, examples of such
dualities are known for cases where the compactification manifolds are
more complicated, such as the manifold $T^{1,1}$
\cite{Klebanov:2000hb}.  While one would expect the \ads\
compactifications studied here to have a 2+1-dimensional CFT dual, we
have few clues as to what the dual theory would be.  The 't Hooft 
coupling is given by the \ads\ scale to be $\lambda\sim |e^{-4iaS}|$, 
but we cannot say much else.  Some clue to the duality
could be found by relating our backgrounds to the 
near-horizon limit of some 2-brane geometry and
the IR limit of its worldvolume theory.  For an NK compactification,
we could perhaps use the fact that a cone with a NK base has holonomy
$G_2$, as suggested in \cite{Behrndt:2004km} in the type IIA context. 
Then our backgrounds would be the near-horizon limit of a
2-brane at the tip of such a $G_2$ cone.  However, the heterotic string
is lacking in 2-branes, so such a picture would likely arise via some
duality. 

Similarly, an important goal to pursue is to relate the many different
types of flux compactifications in the various different theories to
each other, forming a single coherent picture.  Besides relating to
heterotic M-theory solutions, heterotic flux vacua are U-dual to
much-studied flux vacua in IIA \cite{Cardoso:2002hd,
Dall'Agata:2003ir, Behrndt:2004km, Lust:2004ig, DeWolfe:2005uu}, IIB
\cite{Giddings:2001yu, Frey:2004rn, Behrndt:2005bv, Gurrieri:2002iw,
Denef:2005mm}, and M-theory \cite{Gauntlett:2002sc, Behrndt:2003zg,
Kaste:2003zd, Lukas:2004ip, Behrndt:2004bh}.  Though in some examples
these dualities have been made explicit \cite{Becker:2003yv}, the
general connections between all flux vacua have yet to be elucidated.
In addition to dualities, dynamical transitions among vacua are
possible.  As noted above, certain vacua could be related to other by
domain walls or tunneling.  A thorough understanding of these
connections would be a basis for a cartography of the landscape of
flux vacua.

\subsection{Future Directions}\label{ss:future}

In the analysis of section \ref{s:SUSY}, rather than specifying the
gaugino condensate from the outset, we deliberately worked with the
most general case possible.  We found the usual condensate $\Sigma
\sim \Omega + c.c.$ consistent with the NK compactification,
but, in general, as seen from (\ref{SigmaJ}), $\Sigma$ also has $(2,1)$
and $(1,2)$ components.  As discussed in section
\ref{ss:gauginocondensate}, the usual condensation mechanism may
generate these unusual components as a result of the non-K\"ahler
geometry.  However, there may be other, as yet unknown, mechanisms
to generate such condensates, and we see no reason to exlude
them a priori.

Following this type of reasoning, one could further generalize to
consider the condensation of other, more exotic fermion bilinears.
Gravitinos and dilatinos, along with other components of the gaugino,
could conceivably condense through some unknown quantum effect.  One
could simply posit the existence of such a condensate and investigate
its effect on the supergravity solution.  However, we leave such
explorations for future work.  Of course, it will be necessary to 
understand how the 10D 1PI effective action is modified in the presence
of general condensates, just as we have noted above in the relatively
simpler case of gaugino condensation.

One could also investigate whether our \ads\ compactifications can be
lifted to $\mathnormal{dS}_4$.  
Along the lines of \cite{Kachru:2003aw}, one could add
$\overline{NS5}$-branes to break supersymmetry and increase the vacuum energy.
However, one would then need to be sure to stabilize the
$\overline{NS5}$ moduli.

Further, the heterotic flux vacua discussed here should lift to heterotic
M theory in the strong coupling limit.  Heterotic M theory is
attractive for phenomenological model building, and flux
compactifications of eleven dimensions with non-perturbative effects
have been extensively studied \cite{Curio:2001qi}.  
In addition to gaugino condensation on
the $E_8$ branes, open M5 brane instantons are needed to stabilize the
orbifold length.  Furthermore, both \ads\ solutions
\cite{Buchbinder:2003pi} and metastable $\mathnormal{dS}_4$ 
vacua \cite{Becker:2004gw,Buchbinder:2004im} 
have been constructed in the context of heterotic
M theory.  In fact, our backgrounds are the perturbative description of
the \ads\ backgrounds in \cite{Buchbinder:2003pi}, but including the
backreaction of the condensate and $H$ flux.

\section{Conclusion}\label{s:conclusion}

We have presented, from a primarily 10D perspective, a class of supersymmetric
heterotic \ads\ compactifications with both $H$ flux and gaugino
condensation.  The effects combine to fix all the moduli and to yield
a non-complex internal geometry.  In the general case, we
found supersymmetric backgrounds 
at weak coupling and large internal volume with
exponentially large \ads\ radius.  We also showed that proposed super- and
K\"ahler-potentials can reproduce the correct 4D cosmological constant,
although there appear to be subtleties regarding the derivation of the
superpotential via dimensional reduction which is not yet fully 
understood.

To elucidate the 10D geometry of the supersymmetric \ads\ backgrounds,
we used the $G$-structure formalism that has been used extensively for flux
vacua.  To our knowledge, this is its first application 
in the context of gaugino
condensation.  The form of the condensate was left intentionally
general, so as to include the possibility of being generated by
non-standard, possibly 10D, effects.  Furthermore, because we were not
compactifying on a \cy, a  condensate resulting from the standard
4D mechanism does not necessarily take the standard form in any event.

For a particular choice of flux and condensate all the torsion classes
but one vanished, and we found the internal manifold to be 
nearly K\"ahler, which greatly simplified the analysis.  Here the condensate 
took its usual form, the internal and \ads\ sizes were roughly equal, and 
the we solved the Bianchi identity with a nonstandard embedding to give an 
explicit value for $m$.

Despite the progress made here, important issues remain, such as
verifying the equations of motion and solving the Bianchi 
identity in the general 
case.  The generation and effects of exotic gaugino or 
other fermion condensates
pose interesting questions.  More broadly, we have yet to really
explore the connections, via dualities or domain walls, of the
compactifications presented here to all the other extant flux vacua.

\begin{acknowledgments}
We would like the thank Ben Freivogel, Sergei Gukov, 
Shamit Kachru, Liam McAllister, 
Jeremy Michelson, Andrei Micu, 
Hirosi Ooguri, Sergey Prokushkin, Michael Schulz, 
Al Shapere, Alessandro Tomasiello, and Chengang Zhou for helpful discussions. 
The work of ARF was supported by the John A. McCone fellowship in 
theoretical physics at the California Institute of Technology.  The work
of ML was supported by the DOE under contract DE-FG001-00ER45832.
\end{acknowledgments}

%
%

\appendix
\section{Form and Spinor Conventions}\label{a:conventions}

For our index conventions, we take 
upper case Latin for the full ten dimensions, lower case Greek 
for the four Poincar\'e invariant dimensions, and lower case Latin 
for the internal dimensions.
Hats denote tangent space indices.  We work in a signature in which timelike
norms are negative.

Our differential form conventions are as follows:
\bea
\epsilon_{012\cdots (d-1)} &=& +\sqrt{-g}\textnormal{ for $d$ dimensions}
\nonumber\\
T_{[M_1\cdots M_p]} &=& \frac{1}{p!} \left( T_{M_1\cdots M_p} \pm
\textnormal{permutations}\right)\nonumber\\
\left(\star T\right)_{M_1\cdots M_{d-p}}&=& \frac{1}{p!}\epsilon_{M_1\cdots
M_{d-p}}{}^{N_1\cdots N_p}T_{N_1\cdots N_p}\nonumber\\
T&=& \frac{1}{p!} T_{M_1\cdots M_p} dx^{M_1}\cdots dx^{M_p}\ .\label{forms}
\eea
Wedges and exterior derivatives are defined consistently with those 
conventions.  We also use the notation
\be\label{contraction}
(R\contract S)_{N_1\cdots N_q} = \frac{1}{p!} R^{M_1\cdots M_p}
S_{M_1\cdots M_P N_1\cdots N_q} \ ,\ee
which is common in the $G$ structure literature.

Gamma matrices in tangent space have the algebra 
$\{\Gamma^{\hat M},\Gamma^{\hat N}\}=2\eta^{\hat M\hat N}$.  With these
conventions, a Majorana basis is real and symmetric for spacelike indices
and antisymmetric for time.  Gammas can be converted to coordinate indices
with the vielbein.  We define $\Gamma^{M_1\cdots M_p}=\Gamma^{[M_1}\cdots
\Gamma^{M_p]}$.
The chirality is given by
\be
\label{10Dchiral}
\Gamma_{(\widehat{10})}= \Gamma^{\hat 0}\cdots\Gamma^{\hat 9}=
\frac{1}{10!}\epsilon_{M_1\cdots M_{10}}\Gamma^{M_1\cdots M_{10}}\ .
\ee
We can decompose the $\Gamma$ matrices as
\be
\label{decomp}
\Gamma^\mu = \gamma^\mu\otimes 1\ ,\ \ \Gamma^m = \gamma_{(\hat 4)}\otimes
\gamma^m
\ee
with 4D and 6D chirality $\gamma_{(\hat 4)}=-i\gamma^{\hat 0}\cdots
\gamma^{\hat 3}$, $\gamma_{(\hat 6)}= i \gamma^{\hat 4}\cdots\gamma^{\hat 9}$.
The $\gamma^\mu$ have the same symmetry and reality properties as $\Gamma^M$,
while the $\gamma^m$ are imaginary and antisymmetric.

\section{Gamma Matrix and $SU(3)$ Structure Identities}\label{a:identities}

A comprehensive list
of (anti)commutators appears in \cite{Candelas:1984yd}, although there is
at least one typographical error.  It is necessary to replace
\be\label{typo}
[\gamma_{mnp},\gamma^{rst} ]= 2\gamma_{mnp}{}^{rst}-36\delta_{[mn}^{[rs} 
\gamma_{p]}{}^{t]}\ .\ee

Contractions of gamma matrices are given by
\bea
\label{gammacontractioneven}
\gamma^a \gamma_{m_1... m_{2k}} \gamma_a &=& (d-4k) \gamma_{m_1... m_{2k}} \\
\label{gammacontractionodd}
\gamma^a \gamma_{m_1... m_{2k+1}} \gamma_a &=& (4k-d+2) 
\gamma_{m_1... m_{2k+1}} \, .
\eea
Other useful identities are
\bea
\label{gammaidentity1}
\gamma_{mnp} = \frac{i}{6}\gamma_{(\hat 6)} \gamma^{qrs}\epsilon_{mnpqrs}
&,&
\gamma_{mnpq} = \frac{i}{2}\gamma_{(\hat 6)} \gamma^{rs}\epsilon_{mnpqrs} \\
\label{gammaidentity2}
\eta^\dagger\gamma_{mnpqrs}\eta &=& -i\epsilon_{mnpqrs}
\eea
for positive chirality $\eta$.  Using (\ref{gammaidentity1}), we can see
that $\Omega$ as defined in (\ref{su3def}) 
is self-dual, $\star\Omega=i\Omega$.  Self-duality implies
\be
\label{omegacontractions} 
\Omega^{mnp}\Omega_{mnp}=0\ ,\ \ \b\Omega^{mnp}\Omega_{mnp}=48
\ee
(when combined with the relation $\star\epsilon=1$ for the associated volume
form).

The Fierz identities that we use come from expanding in terms of the complete
set of $\gamma$ matrices.  Specifically, we find
\bea
\label{fierz1}
\eta\eta^\dagger &=& \frac{1}{8}-\frac{i}{16}J_{mn}\gamma^{mn}-
\frac{i}{16}J_{mn}\gamma^{mn}\gamma_{(\hat 6)} +\frac{1}{8}\gamma_{(\hat 6)} \\
\label{fierz2}
\eta\eta^T &=& -\frac{1}{48} \Omega_{mnp}\gamma^{mnp}
\eea
for the normalized positive chirality spinor $\chi$ used in the text.
This identity can be used to show that $J_m{}^nJ_n{}^p=-\delta_m^p$ and also
that
\be
\label{Jdual}
(J\wedge J)_{mnpq}=6J_{[mn}J_{pq]} = \epsilon_{mnpq}{}^{rs}J_{rs}
=2(\star J)_{mnpq}\ .
\ee
Other helpful identities which follow from self-duality of $\Omega$ and
the Fierz identities are
\be\label{omegacontractions2}
\Omega^{mnr}\b\Omega_{pqr} = 4\delta_{[pq]}^{mn}-4J^m{}_{[p}J^n{}_{q]}
+8i\delta^{[m}_{[p}J_{q]}{}^{n]}\ ,\ \ 
\Omega^{mnr}\Omega_{pqr} = 0\ .\ee
The first of (\ref{omegacontractions2}) is also given in 
\cite{Behrndt:2005bv}.

We can also decompose any tensor with respect to the $SU(3)$ structure.
We write a real 3-form $R$ and complex 4-form $S$ as
\bea
R&=& \frac{3i}{2}\im\left(\b R_1\Omega\right) +R_3 +J\wedge R_4\nonumber\\
S&=& S_1 J\wedge J +J\wedge S_2 +\Omega\wedge S_5 \ .
\label{decompose}\eea
We have labeled the components in a fashion consistent with the torsion
modules $W_i$ in equations (\ref{dJtorsion},\ref{dOmegatorsion}).
Then we can invert (\ref{decompose}) to get
\bea
R_1&=& -\frac{i}{6}\Omega \contract R\ ,\ R_{4,p}
= \frac{1}{2} (J\contract R)_p \ ,\nonumber\\
S_1 &=& \frac{1}{12}  (J\wedge J)\contract S\ ,\ 
S_{5,p}= \frac{1}{24} \left(\bar{\Omega}\contract S\right)_p
\label{extraction}
\eea
as in \cite{Dall'Agata:2003ir, Gauntlett:2003cy, Chiossi:2002}.
$R_3$ and $S_2$ are primitive in the sense that $J\contract R_3=0$ and
$J\contract S_2 = 0$.

\section{Contorsion and Intrinsic Torsion}\label{a:torsions}

Here we present a brief review of various definitions and identities 
involving torsions.  These formulae and conventions can be found, for
example, in \cite{Nakahara:1990th,Carroll:2004st}.

The difference between a torsional connection $\b\Gamma$ and
the torsion-free Levi-Civita connection $\Gamma$ is the contorsion tensor
\be
\label{contorsion}
\b\Gamma^m{}_{np} -\Gamma^m_{np}= \kappa^m{}_{np}\ ,\ \
\kappa_{mnp}=-\kappa_{pnm}\ ,  
\ee
where the antisymmetry follows from metric compatibility.  Then, because
the vielbein must be covariantly constant with respect both to the 
torsionful and torsionless derivatives $\b\Del,\Del$, the spin connection 
is shifted by
\be\label{contorsion2}
\b\omega_m{}^{\hat a}{}_{\hat b} -\omega_m{}^{\hat a}{}_{\hat b} 
= \kappa^p{}_{mn}e^{\hat a}_p e^n_{\hat b}
\ \ \left( = -\kappa_m{}^{\hat a}{}_{\hat b} 
\textnormal{ for $\kappa$ totally antisymmetric}
\frac{}{}\right)\ .\ee
The Riemann tensor is still given by the usual formulae
\be\label{riemanntorsion}
\b R^m{}_{npq} = 2\del_{[p}\b\Gamma^m{}_{q]n}+2\b\Gamma^m{}_{[p|r|}
\b\Gamma^r{}_{q]n}\ ,\ \ 
\b R^{\hat a}{}_{\hat b} = d\b\omega^{\hat a}{}_{\hat b} 
+\b\omega^{\hat a}{}_{\hat c}\wedge\b\omega^{\hat c}{}_{\hat b}\ .\ee

The (intrinsic) torsion $\tau$ is
defined by  $\b\Del_{[n}\b\Del_{p]} f= -(1/2)\tau^m{}_{np}\b\Del_m f$ for a
scalar $f$
and is related to the contorsion by
$2\kappa^m_{[np]}=\tau^m{}_{np}$.  The torsion $\tau$ is totally 
antisymmetric and modifies the usual relations
\be\label{commutators}
\left[\b\Del_{m},\b\Del_{n}\right] v^p 
= \b R^p{}_{qmn}v^q -\tau^q{}_{mn}\b\Del_q v^p \ ,\ \ 
\left[\b\Del_{m},\b\Del_{n}\right]\psi = \frac{1}{4}\b R_{mnpq}\gamma^{pq}
\psi\ee
for vectors and spinors.
Also, the torsion gives a topological obstruction to finding special 
holonomy with the Levi-Civita connection, as reviewed in section 
\ref{ss:difftorsion}.

\section{Supergravity Potential Normalizations}\label{a:normalization}

Here we will describe the normalization of the 4D $\N=1$ supergravity
variables in the effective field theory description.  We roughly follow
\cite{Denef:2005mm}.  After dimensional reduction on the metric 
$g_{mn}=e^{2u}\t g_{mn}$, we find a 4D string (or Jordan) frame action
\be\label{jordanaction}
S=\frac{\t V_6}{(2\pi)^7\ap{}^4}\int d^4 x\sqrt{-g} e^{6u-2\phi}R(g)+\cdots
\ ,\ee
where $\t g_{mn}$ has volume $\t V_6$, and we have used the correct string
theory value for the 10D gravitational coupling.  Rescaling to Einstein
frame $g_{\mu\nu}=e^{2\phi-6u}g_{E,\mu\nu}$, we find
\be\label{einsteinaction}
S= \int d^4x \sqrt{-g_E} \left( \frac{m_p^2}{2}R(g_E)-V+\cdots\right)\, ,\ \
\frac{m_p^2}{2} = \frac{\t V_6}{(2\pi)^7\ap{}^4}\ .\ee
We have now included the $\N=1$ supergravity potential, witten in terms of the
superpotential and K\"ahler potential as
\be\label{sugraV}
V=\frac{1}{m_p^2}e^{\mathcal{K}}\left( \mathcal{K}^{i\bj}D_i W D_{\bj}\b W
-3|W|^2\right)\ ,\ee
where we can write
$\Lambda_E =-V/m_p^2$ for the absolute
Einstein frame cosmological constant in an \ads\ vacuum.  Note that
this sets our conventions for the cosmological constant, as well.
Henceforth, we will take $\t V_6=(2\pi\sqrt{\ap})^6$, as in the text, though
the normalization can be generalized.  In this appendix, we are 
approximating the warp factor as trivial and the dilaton as constant over
$X^6$.

We take the heterotic superpotential of 
\cite{Becker:2003gq,Cardoso:2003af,Cardoso:2003sp}
(generalizing that of \cite{Gukov:1999ya}) to be normalized as
\be\label{gvwsuper}
W= \frac{m_p^3}{\sqrt{4\pi}} \frac{1}{(2\pi\sqrt{\ap})^5}
\int (H+ibdJ+c\Sigma)\wedge\t\Omega\ .\ee
The factors of $2\pi\sqrt{\ap}$ remove the dimensionality of the integral,
so that only the 4D Planck scale enters the superpotential as a 
dimensional factor.  The relative normalizations $b,c$ of the torsion and
condensate terms are addressed in section \ref{ss:potentials}.
As discussed in section \ref{ss:potentials}, we use the K\"ahler potential 
\be\label{kahlerpot}
\mathcal{K}=-3\ln(-i(T-\b T)) -\ln(-i(S-\b S))
-\ln\left(\frac{i}{(2\pi\sqrt{\ap})^6} \int\t\Omega\wedge\b{\t\Omega}\right)
+\delta\mathcal{K}\ ,\ee
where we have included a constant $\delta\mathcal{K}$ in order to fix
the potential given a normalization of $W$.   In terms of the 10D variables, the 4D moduli are $\im\, T = e^{2u}$, $\im\, S =e^{6u-2\phi}$ for the heterotic theory.\footnote{The real parts of $S$ and $T$ are axions related to $B_{\mu\nu}$ and $B_{mn}$, respectively.}  We are ignoring warping and also variation of
the dilaton in the compact space.

So fix $\delta\mathcal{K}$, we consider the tension of a BPS domain wall,
which is given by the jump in superpotential over the wall,
$T=2e^{\mathcal{K}/2}|\Delta W|$.  If we take a \cy\ compactification,
an NS5-brane on a SLAG 3-cycle $c$ is a BPS domain wall.  Crossing the domain 
wall, the flux jumps one unit on the dual cycle, $\int_{\t c} \Delta H =
4\pi^2\ap$ according to the Dirac quantization condition.  Since 
$c$ is calibrated, we find
\be\label{normalK1}
|\Delta W| = \frac{m_p^3}{\sqrt{4\pi}} \frac{\t V_c}{(2\pi\sqrt{\ap})^3}\ ,\ee
where $\t V_c$ is the volume of $c$ with respect to $\t g_{mn}$. The 
domain wall tension is then
\be\label{normalK2}
T = \frac{\t V_c}{\sqrt{8}}
\frac{1}{(2\pi)^5\ap{}^3}e^{\phi-6u}e^{\delta\mathcal{K}/2}\ .\ee
Comparing to the Einstein-frame action of an NS5-brane wrapping $c$, we
find $\delta\mathcal{K}=3\ln 2$.


\bibliography{adssusyvac}

\end{document}